\documentclass[12pt]{article}
\usepackage[figuresright]{rotating} 
\usepackage{inputenc}
\usepackage{graphicx,subcaption}
\usepackage{url}
\usepackage{footnote}
 \usepackage[flushleft]{threeparttable}
\usepackage[top=1.25in, bottom=1.25in, left=1in, right=1in]{geometry}
\usepackage{microtype}

\usepackage[english]{babel}
\newcommand{\sizea}{\fontsize{10.5pt}{12.6pt}\selectfont} 

\usepackage{titlesec}
\titlespacing\section{0pt}{12pt}{0pt}
\titlespacing\subsection{0pt}{6pt}{0pt}
\titlespacing\subsubsection{0pt}{6pt}{0pt}

\titlespacing\section{0pt}{10pt}{10pt} 
\titlespacing\subsection{0pt}{8pt}{8pt} 

\usepackage{makecell}
\usepackage{booktabs}
\usepackage[table]{xcolor}
\usepackage{float}
\usepackage{caption}
\usepackage[sort]{natbib}

\usepackage{changepage} 
\usepackage{times}

\usepackage{afterpage}
\usepackage{pdflscape}

\usepackage{xcolor}
\usepackage{amsmath}
\usepackage{amsthm}
\usepackage{amssymb,bbm,wasysym,tikz}
\usepackage{lscape}
\usepackage{verbatim}
\usepackage{enumitem}
\usepackage{booktabs}
\usepackage{multicol}
\usepackage{multirow}
\usepackage{longtable,booktabs} 
\usepackage{setspace}
\usepackage[normalem]{ulem} 

\usepackage{xfrac}

\usepackage{multibib}
\newcites{app}{References}
\usepackage[bottom]{footmisc}

\usepackage{wasysym}
\usepackage{tikz}
\newcommand{\none}{\tikz\draw[fill=black!0] (0,0) circle (0.4em);} 
\newcommand{\weak}{\tikz\draw[fill=black!10] (0,0) circle (0.4em);} 
\newcommand{\medium}{\tikz\draw[fill=black!45] (0,0) circle (0.4em);} 
\newcommand{\high}{\tikz\draw[fill=black!100] (0,0) circle (0.4em);}
\newcommand{\sym}[1]{\ifmmode^{#1}\else\(^{#1}\)\fi}

\usepackage{array}
\newcommand{\PreserveBackslash}[1]{\let\temp=\\#1\let\\=\temp}
\newcolumntype{C}[1]{>{\PreserveBackslash\centering}p{#1}}
\newcolumntype{R}[1]{>{\PreserveBackslash\raggedleft}p{#1}}
\newcolumntype{L}[1]{>{\PreserveBackslash\raggedright}p{#1}}

\definecolor{mypink1}{RGB}{255, 204, 255}
\definecolor{mypurple1}{RGB}{204,51,255}
\definecolor{mygreen1}{RGB}{61,245,0}
\definecolor{myorange1}{RGB}{153,76,0}
\definecolor{myred}{RGB}{121,36,35}
\newcommand*{\rom}[1]{\expandafter\@slowromancap\romannumeral #1@}
\usepackage{hyperref}

\hypersetup{
	colorlinks   = true, 
	urlcolor     = blue, 
	linkcolor    = blue, 
	citecolor   = blue  
}

\newcommand{\ra}{\uppercase\expandafter{\romannumeral 1}}
\newcommand{\rb}{\uppercase\expandafter{\romannumeral 2}}
\newcommand{\rc}{\uppercase\expandafter{\romannumeral 3}}

\newcommand{\hide}[1]{}

\newtheorem{result}{Result}

\theoremstyle{plain}

\theoremstyle{definition}

\setcitestyle{citesep={;}}

\makeatletter
\def\@fnsymbol#1{\ensuremath{\ifcase#1\or\relax\fi}}
\makeatother

\begin{document}
\begin{singlespace}
\title{How General Are Measures of Choice Consistency? \\ Evidence from Experimental and Scanner Data}
\author{
\vspace{-0.2cm}
\setstretch{1.5} 
    \begin{tabular}{>{\centering\arraybackslash}p{9em} >{\centering\arraybackslash}p{9em} >{\centering\arraybackslash}p{9em}}
        Mingshi Chen & Tracy Xiao Liu & You Shan \\
        Shu Wang & Songfa Zhong & Yanju Zhou\thanks{
        \hspace*{-1cm}
        \begin{minipage}[t]{1.05\textwidth}
        \hspace*{1em} Chen, Liu and Wang: School of Economics and Management, Tsinghua University (Chen: cms19@mails.tsinghua.edu.cn; Liu: liuxiao@sem.tsinghua.edu.cn; Wang: shu-wang20@mails.tsinghua.edu.cn).
        Shan: Faculty of Business for Science and Technology, School of Management, University of Science and Technology of China (shanyou@ustc.edu.cn). Zhong: Department of Economics, National University of Singapore (zhongsongfa@gmail.com). Zhou: School of Business, Central South University (zyj4258@sina.com). We thank David Ahn, Syngjoo Choi, Federico Echenique, Shachar Kariv, Yusufcan Masatlioglu, Bin Miao, John Quah, Satoru Takahashi, seminar participants at University of Michigan, University of Maryland, Rutgers University, Singapore Management University, Lingnan University, Nankai University, and participants at Bounded Rationality in Choice Conference (BRIC-X, 2024), Decision: Theory, Experiments, and Applications (D-TEA, 2024) and Foundations of Utility and Risk (FUR, 2024). Liu gratefully acknowledges financial support by NSFC (72222005, 72342032). Our study was approved by National University of Singapore Institutional Review Board (ECSDERC-2021-7) and was registered in the AEA Registry under trials AEARCTR-0007467 and AEARCTR-0008733.\\
        \end{minipage}}
    \end{tabular}
}

\maketitle

\vspace{-1.3cm}

\begin{abstract}
Choice consistency with utility maximization, as a key assumption in economics, has been extensively used to evaluate decision quality of individuals and to predict real-world outcomes across different contexts. Here we investigate the generalizability of consistency measures derived from budgetary decisions in the lab-in-the-field experiment and purchasing decisions using supermarket scanner data. In the first study, we observe a lack of correlation between consistency scores derived from risky decisions in the experiment and those from supermarket food purchasing decisions. In the second study, we observe moderate correlations between experimental tasks and low to moderate correlations across purchasing categories and over time periods within the supermarket. Moreover, consistency in the two settings exhibits distinct predictive validity in predicting consumer behavior. These results suggest that choice consistency, as a measure of decision quality, may be better characterized as a multidimensional skill set rather than a single-dimensional ability.

\end{abstract}

\end{singlespace}
\noindent \text{Keywords}: Consumer choice, utility maximization, revealed preference, lab-in-the-field experiment

\noindent \text{JEL}: C91, D81, D91 

\clearpage
\section{Introduction}
\label{sec:introduction}

Central to economic analysis is the assumption that a decision maker (DM) maximizes her utility function given her budget constraint. Revealed preference analysis characterizes the conditions under which a DM, based on a given choice dataset, indeed maximizes a well-behaved utility function \citep{samuelson1938note,afriat1967construction,varian1982nonparametric}. It thus provides a powerful framework and serves as a textbook example of normative approaches on how decisions should be made. Going beyond normative approach, researchers have examined the descriptive validity of this framework, and used it to assess choice consistency in various settings \citep{crawford2014empirical,chambers2016revealed}. A growing body of literature has measured consistency scores in risky, intertemporal, and social budgetary decisions in experiments, as well as in purchasing decisions derived from expenditure surveys and from scanner data in supermarkets.\footnote{For studies utilizing experimental data, see \citet{ahn2014estimating,andreoni2002giving,choi2007consistency,echenique2023approximate,fisman2007individual,halevy2018parametric,polisson2020revealed,dembo2024ever,ellis2024revealing}, and for studies employing purchasing data, see \citet{blundell2003nonparametric,blundell2008best,crawford2010habits,cherchye2020revealed,dean2016measuring,echenique2011money}. Consistency with utility maximization is often referred to as choice consistency or rationality. For simplicity, we use choice consistency throughout this paper.} 

Furthermore, it has been suggested that consistency scores reflect an individual's capacity to make sound decisions. This notion has been proposed as a significant factor contributing to wealth disparities among individuals in the general population \citep{choi2014more} and development gaps across countries \citep{cappelen2023development}. Choice consistency has also been used to assess the impact of financial constraints due to payday and loan \citep{carvalho2016poverty,carvalho2024misfortune}, and to evaluate the effectiveness of policy interventions in education and poverty \citep{kim2018role,li2023multifaceted}.

Building on these insights into choice consistency and its applicability across settings, we examine its empirical validity and test its generalizability---the extent to which DMs with high consistency scores in one setting also demonstrate high scores in another. More specifically, do DMs with high consistency scores as consumers in a supermarket environment also have higher scores as participants in a controlled experimental setting? Two related questions arise. Do DMs show correlated consistency scores across risk, social, and food-related decisions in the experiment? Do DMs exhibit correlated consistency scores across consumption categories and time periods in the supermarket? On the one hand, it is plausible that measures of choice consistency can be generalized across settings, as they may reflect a DM's overall capacity of making good decisions. Conversely, it is also conceivable that these measures are setting-specific, as DM may possess different preference structures, face varying budget constraints, and have distinct experiences or decision rules across settings, leading to divergent scores inferred from choice data.

We address these questions by combining experimental and scanner data. First, we examine consistency scores of the same individuals who make risky decisions in the lab-in-the-field experiment and make consumption decisions in the supermarket. Using the scanner data, we restrict our attention to 6,144 consumers of whom we have purchase records of meat and vegetables—the two most common consumption categories—over 24 consecutive months, so that we have sufficient power for the revealed preference analysis (see Section~\ref{sec:design}). We then successfully invite 1,055 of these 6,144 consumers to participate in a budgetary task (Experiment 1) in which participants make 22 portfolio decisions between two Arrow securities with different budget lines \citep{choi2007consistency,choi2014more,halevy2018parametric,kim2018role}. They need to allocate 100 experimental tokens between two accounts in which tokens are converted to cash with different exchange rates, and the amount in one of the two accounts will be paid out randomly with 50 percent probability. Afterwards, by extracting these 1,055 consumers from the scanner data and aggregating the consumption at the month level, we obtain the choice dataset in the supermarket (Scanner Dataset~1).

We measure the choice consistency by evaluating the extent to which choices adhere to the Generalized Axiom of Revealed Preference (GARP), a necessary and sufficient condition whereby a dataset can be rationalized by a well-behaved utility function (in accordance with utility maximization). To assess how closely an individual's choice dataset complies with GARP, we compute the commonly used critical cost efficiency index \citep[CCEI,][]{afriat1972efficiency}, as well as the Houtman-Maks index \citep[HMI,][]{houtman1985determining}; money pump index \citep[MPI,][]{echenique2011money}; and minimum cost index \citep[MCI,][]{dean2016measuring}.
Based on Experiment~1 and Scanner Dataset~1, our primary finding is that consistency scores are uncorrelated between risky decisions in the experiment and food consumption decisions in the supermarket regardless of the index used. While this observation might appear to be unsurprising to some, it does contrast with our informal expert prediction.\footnote{One of the authors conducted an informal expert prediction when presenting this paper at the Bounded Rationality in Choice Conference (BRIC, 2024, Venice) and the Decision: Theory, Experiments, and Applications Workshop (D-TEA, 2024, Paris). Among the approximately 50 attendees at each conference, majority predicted moderate to high correlations between the consistency scores of risky decisions and food decisions, while only 2 to 3 attendees anticipated the correlation to be close to zero.}

Second, we test whether the observed near-zero correlation is attributed to differences in the types of decision---risky decisions versus food consumption decisions---or due to the disparities between experimental and supermarket settings and conduct a second experiment (Experiment 2). We invite another 302 consumers to participate in three budgetary tasks, including the risk task used in Experiment 1, the social task \citep{andreoni2002giving,fisman2007individual} and the food task \citep{chen2023emergence,harbaugh2001garp}. We find that consistency scores are moderately correlated across these three tasks, where the magnitude of the correlations is the highest between risk and social tasks.

Third, we also examine the correlation of consistency scores of choices within the supermarket using Scanner Dataset 2. The dataset consists of 822 consumers in different consumption categories, including meat, vegetables, fruits, and snacks, and 938 consumers in different time periods from 2018 to 2021. We find that the correlations of consistency scores are generally significantly positive between consumption categories and over different time periods within the supermarket, although the magnitude is relatively low.

We conduct several robustness checks on the observed correlations. Specifically, we (1) utilize alternative indices for HMI, MPI, and MCI; (2) control for the power to detect GARP violations; (3) impose additional assumptions on preference structures through revealed preference analysis; and (4) assess the fit of preference structural estimations. Our findings remain consistent across these variations, with two notable exceptions. Namely, when we impose more assumptions on preference structures or evaluate the fit of structural estimations, the correlations within the scanner data significantly improve, both across categories and over time.

Our results indicate that choice consistency exhibits multidimensional characteristics, with distinct factors affecting the degree of consistency measures across different decision environments, such as randomness in choice behaviors, budget constraints, formation of preferences, and the application of heuristic rules. In particular, we find that the choices made in supermarkets are more likely to come from random DMs compared to those made in experimental settings. Furthermore, the consistency of supermarket choices is influenced by various contextual factors, such as seasonality, time of day, and individual shopping experiences. In contrast, consistency in experimental choices is shaped by learning effects across rounds of decisions, the use of heuristic rules, and participants' cognitive abilities. These distinct elements play crucial roles in how individuals navigate decision-making in the supermarket and in the experiment. To support this distinction, we further show that consistency in experimental data predicts the frequency and magnitude of discount usage in consumer's purchasing choices, while consistency in supermarket data predicts the month-to-month volatility in shopping patterns. Taken together, we highlight the multifaceted nature of decision-making across settings.

Our study adds to the literature on measuring consistency with utility maximization based on revealed preference analysis. In addition to theoretical and experimental studies, choice consistency measures have been proposed as proxies of decision-making quality and have been widely used in applied settings. These studies apply choice consistency measures to examine individual differences in wealth, occupation, and borrowing behavior of high-cost loans, policy interventions aimed at education and poverty reduction, as well as country-level variations in development \citep[e.g.,][]{banks2019education,cappelen2023development,carvalho2016poverty,carvalho2024misfortune,choi2014more,kim2018role,li2023multifaceted}.
In the literature, most studies focus on a specific type of decision in either the experiment or the field.\footnote{An exception is \cite{kim2018role}, in which they collect choice data on both risky and intertemporal decisions in the laboratory experiment to evaluate the effects of an education program. While their study does not examine the correlation between consistency scores of risky and intertemporal decisions, we use their data and compute the Spearman's rank correlation to be 0.51, which is in line with the observed correlations across tasks in our Experiment~2.} Our study is the first to examine the generalizability of consistency measures across settings, and contributes to a better understanding of the concept of utility maximization.

Our study also contributes to the literature on the generalizability of choice behaviors across settings \citep{camerer2015promise,levitt2007laboratory, snowberg2021testing,chapman2024looming}. A question that has attracted great attention is the generalizability of risk preferences. For example, \cite{weber2002domain} show that risk attitudes vary between domains; these include financial, health/safety, recreational, ethical, and social decisions.
\cite{dohmen2011individual} find that different measures of risk attitudes are imperfectly correlated in different settings. \cite{barseghyan2011risk} use three deductible choices made in the domain of auto and homeowner insurance to separately estimate the individual risk attitude using a structural approach, and reject the null of fully domain-general risk attitudes. \cite{einav2012general} provide evidence to support positive correlations of risk preferences within five employer-provider insurance coverage decisions, and weaker relationships between these insurance decisions and investment decisions with respect to a 401(k) plan. In addition to risk preferences, moderate or weak correlations between settings have also been reported in time preferences \citep{,augenblick2015working,burks2012measures}; social preferences \citep{bruhin2019many,fehr2011field,galizzi2019external}; and strategic sophistication \citep{georganas2015persistence,levitt2011checkmate,rubinstein2016typology}. These studies have made important contributions to the extent to which theory-guided experimental findings can be generalized and applied to real-world scenarios, which is essential for theoretical development, experimental design, and practical application.

In this study, we examine choice consistency using risky, social, and food decisions in two experiments and consumption purchases in the supermarket. On the one hand, the lack of correlation between experimental and scanner data suggests that consistency scores revealed from choice behaviors in various settings may not solely reflect the general ability to make good decisions according to one’s preferences. On the other hand, we do observe moderate correlations across three tasks in the experiment and some weak to moderate correlations across categories and time in the supermarket, which indicate that consistency measures can be generalized within settings. 

In addition to measurement and identification issues, these findings support that choice consistency is multidimensional. This perspective resonates with developments in the literature on intelligence, which have evolved over decades from a one-dimensional understanding of intelligence \citep{spearman1904general} to the recognition of two dimensions: fluid and crystallized intelligence \citep{cattell1943measurement}, and ultimately to the widely accepted view of multidimensional intelligence \citep{mcgrew2009chc,horn1966refinement}. In the economics literature, accumulative evidence has provided support for the notion that choice behaviors are interconnected and can be reduced to several underlying common factors \citep{chapman2023econographics, dean2019empirical,stango2023we}, and the ability to make good decisions is multidimensional \citep{deming2021growing,cerigioni2021dual,ilut2023economic}. Building on these studies, our study takes an initial step to address an important yet unexplored question on the generalizability of choice consistency, a central notion in economic analysis. 
 
The remainder of the paper is organized as follows. Section~\ref{sec:theory} presents the theoretical background of revealed preference analysis. Section~\ref{sec:design} describes the scanner data and the experimental design. Section~\ref{sec:results} presents our main results, and Section~\ref{sec:mechanism} examines the underlying mechanisms. Section~\ref{sec:validity} investigates the predictive validity of consistency scores across settings. Section~\ref{sec:conclusion} provides some concluding remarks.

\section{Theoretical Background}
\label{sec:theory}

Consider a DM who chooses bundles $x^{t}\in \mathbb{R}_{+}^{K}$
from budget lines \{$x:p^{t}\cdot x\leq p^{t}\cdot x^{t},p^{t}\in \mathbb{R}%
_{++}^{K}$\}. A \textit{dataset} $\mathcal{O}=\left\{ \left(
p^{t},x^{t}\right) \right\} _{t=1}^{T}$ refers to a collection of
$T$ decisions of the DM. Let $%
\mathcal{X}=\{x^{t}\}_{t=1}^{T}$ be the set of bundles chosen by the DM. We say that $x^{i}$ is \emph{directly revealed preferred} to 
$x^{j},$ denoted by $x^{i}\succsim ^{\ast }x^{j},$ if a DM chooses $%
x^{i}$ when $x^{j}\in \mathcal{X}$ is affordable (i.e., $p^{i}\cdot
x^{j}\leq p^{i}\cdot x^{i}$). Denote the asymmetric part as $\succ ^{\ast },$
which refers to the relation of \emph{directly strictly revealed preference.}
Denote $\succsim ^{\ast \ast }$ the transitive closure of $\succsim ^{\ast },
$ which refers to the \emph{revealed preferred} relation. A dataset $%
\mathcal{O}$ satisfies the Generalized Axiom of Revealed Preference (GARP) if the following holds: 
\begin{equation}
\text{for all }x^{i}\text{ and }x^{j},\text{ }x^{i}\succsim ^{\ast \ast
}x^{j}\text{ implies }x^{j}\nsucc ^{\ast }x^{i}.  \label{garp}
\end{equation}

We say that a utility function $U:\mathbb{R}%
_{+}^{K}\rightarrow \mathbb{R}$ \emph{rationalizes} the dataset $\mathcal{O}
$ if for every bundle $x^{t}:$ 
\begin{equation*}
U(x^{t})\geq U(x)\text{ for all }x\in \mathbb{R}_{+}^{K}\text{ s.t. }%
p^{t}\cdot x\leq p^{t}\cdot x^{t}.
\end{equation*}%
Afriat's theorem %
\citep{afriat1967construction,varian1982nonparametric} states that a
dataset can be rationalized by some well-behaved (continuous and strictly increasing) utility function, if and only if the dataset obeys GARP.

\subsection{Indices of Choice Consistency}

Afriat's theorem informs whether a dataset can be rationalized by some utility function. When GARP is satisfied, we know that the dataset is consistent with utility maximization; however, when GARP is violated, it does not provide information about the extent of inconsistency. To address this, several indices have been developed to quantify the degree of consistency in the dataset. Below we briefly review four of them.

\textbf{Critical Cost Efficiency Index (CCEI)} \ A popular approach to measure the departure from the maximization of utility is the \emph{critical cost efficiency index} (CCEI) proposed by \cite{afriat1972efficiency}.
A DM has a CCEI $e\in \left[ 0,1\right] $ if $e$ is the largest number with a well-behaved $U$ that rationalizes the dataset for every $%
x^{t}\in \mathcal{X}:$ 
\begin{equation}
U(x^{t})\geq U(x)\text{ for all }x\in \mathbb{R}_{+}^{K}\text{ s.t. }%
p^{t}\cdot x\leq ep^{t}\cdot x^{t}.  \label{garpe}
\end{equation}%
A CCEI of 1 indicates passing GARP perfectly. A CCEI less than 1---say 0.95---indicates that there is a utility function for which the chosen bundle $x^{t}$ is preferred to any bundle that is more than 5 percent cheaper than $x^{t}$. Put differently, the CCEI can be viewed as the amount by which a budget constraint must be relaxed to remove all GARP violations, because the DM is spending more money than is required to achieve the utility targets \citep{polisson2024rationalizability}. Alternatively, it may be related to the behavioral notion of just-noticeable difference, the innate inability of DM to distinguish similar
bundles \citep{dziewulski2020just}. 

\textbf{Houtman-Maks Index (HMI)} \ \citet{houtman1985determining} propose an alternative approach by measuring the maximum number of observations in the observed sample that are consistent with rational choices. For example, an HMI score of 0.941 indicates that the highest proportion of subset of choices consistent with GARP is 94.1 percent.

\textbf{Money Pump Index (MPI)} \ \citet{echenique2011money} provide an index to measure the amount of money one can extract from an individual for each violation of GARP. Their index is based on the idea that an individual with a GARP violation can be exploited by an ``arbitrager'' as a ``money pump''. The ``arbitrager'' can choose the allocation $x^1$ at price $p^2$ and the allocation $x^2$ at price $p^1$, then trade $x^1$ with the individual at $p^1$ and $x^2$ at $p^2$, which yields a profit of $p^1(x^1-x^2)+p^2(x^2-x^1)$. Given multiple violations of GARP, a money pump cost will be associated with each violation. Following \citet{echenique2011money}, we use the mean money pump costs for cyclic sequences of allocations with the length of two. For example, an MPI score equal to 0.059 means on average 5.9 percent of expenditure can be exploited by an ``arbitrager" from GARP violation.

\textbf{Minimum Cost Index (MCI)} \ \citet{dean2016measuring} propose MCI to measure the minimum cost of breaking all revealed preference cycles in a dataset. MCI is defined as $\min\limits_{B\subset R_0}\frac{\sum_{(i,j)\in B}p^i(q^i-q^j)}{\sum_{t=1}^T p^t q^t}$ such that $R_0/B$ is acyclic, where $R_0$ represents the set of the preference relation $\succsim ^{\ast \ast }$.
The index has a high value when there are a large number of cycles for which all GARP violations are based on significant monetary differences relative to total expenditure. Thus, the MCI responds to both the number and severity of revealed preference violations. For example, if the MCI is 0.004, it means that the average cost of preference relations that must be removed to render the dataset consistent with GARP is 0.4 percent. 

In summary, these indices measure the degree of GARP violations from different perspectives. More detailed theoretical discussions can be found in \citet{apesteguia2015measure}, \citet{echenique2021meaning} and \citet{polisson2024rationalizability}.

\section{Experimental Design and Scanner Data}
\label{sec:design}

In this section, we describe the experimental design and the scanner data. Our analysis focuses on consumers who shop at a leading retail company with more than 400 supermarkets spread over six provinces in southern China. We conducted two lab-in-the-field experiments using these consumers to measure consistency scores for risky, social, and food decisions. The combination of experimental and scanner datasets enables us to test the generalizability of measures of choice consistency.

\subsection{Experiments}
\label{sec:risk}
In Experiment 1, we measure consistency scores for risky decisions using a standard budgetary design \citep{choi2007consistency}. In each round of the task, participants allocate 100 tokens between two contingent assets (accounts) and receive the money in one asset with 50 percent probability. The exchange rate between token and cash differs for the two assets and varies across decisions. If participants are risk neutral or risk seeking, they would allocate all tokens to the cheaper asset; if they are risk averse, they would allocate some tokens to each asset depending on their risk preferences.

\begin{figure}[hbtp]
    \centering
\begin{subfigure}[b]{0.32\textwidth}

        \includegraphics[width=1\linewidth]{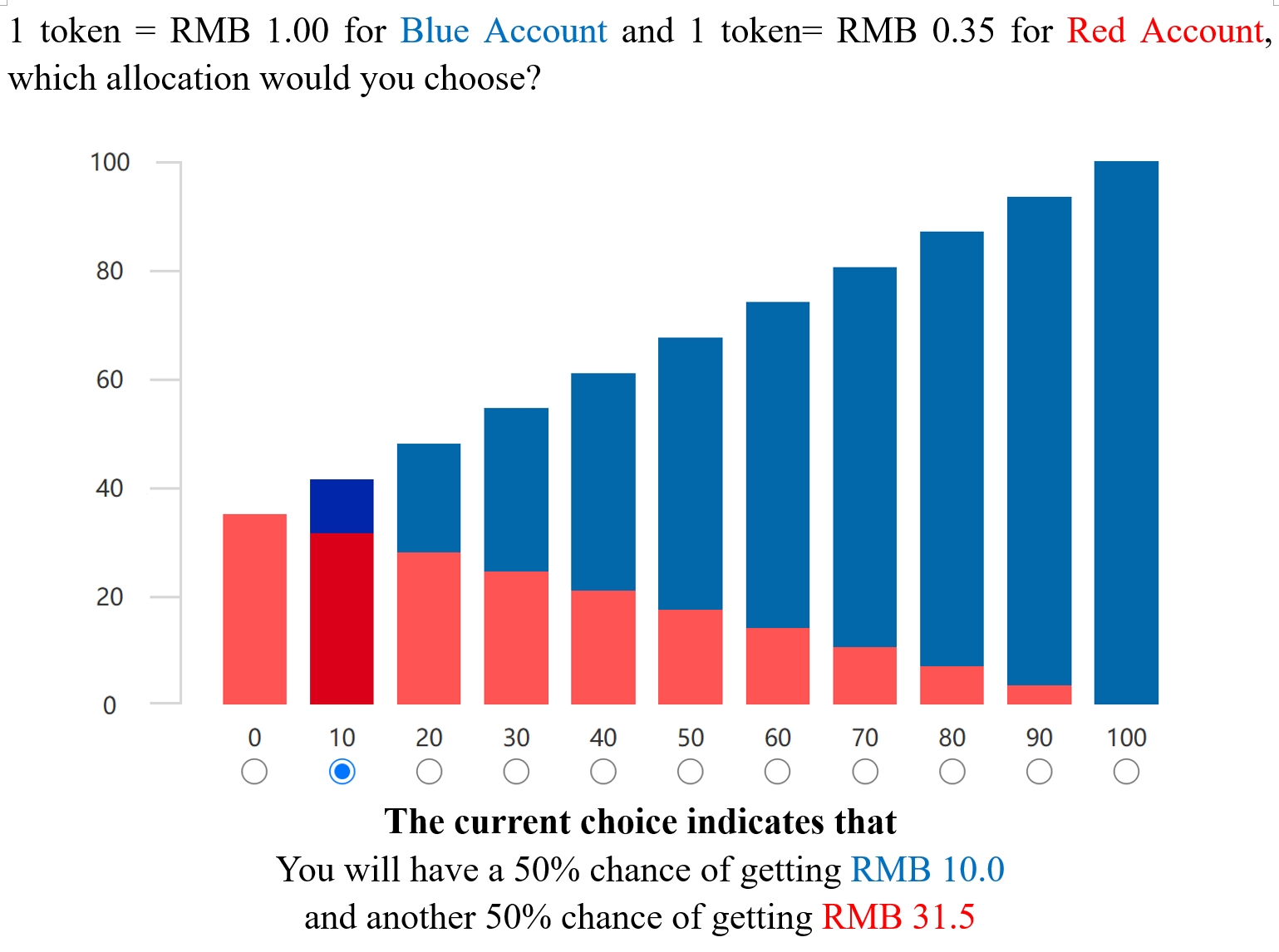}
        \caption{Risk Task}
        \label{fig:risk}

\end{subfigure}
\begin{subfigure}[b]{0.32\textwidth}

        \includegraphics[width=1\linewidth]{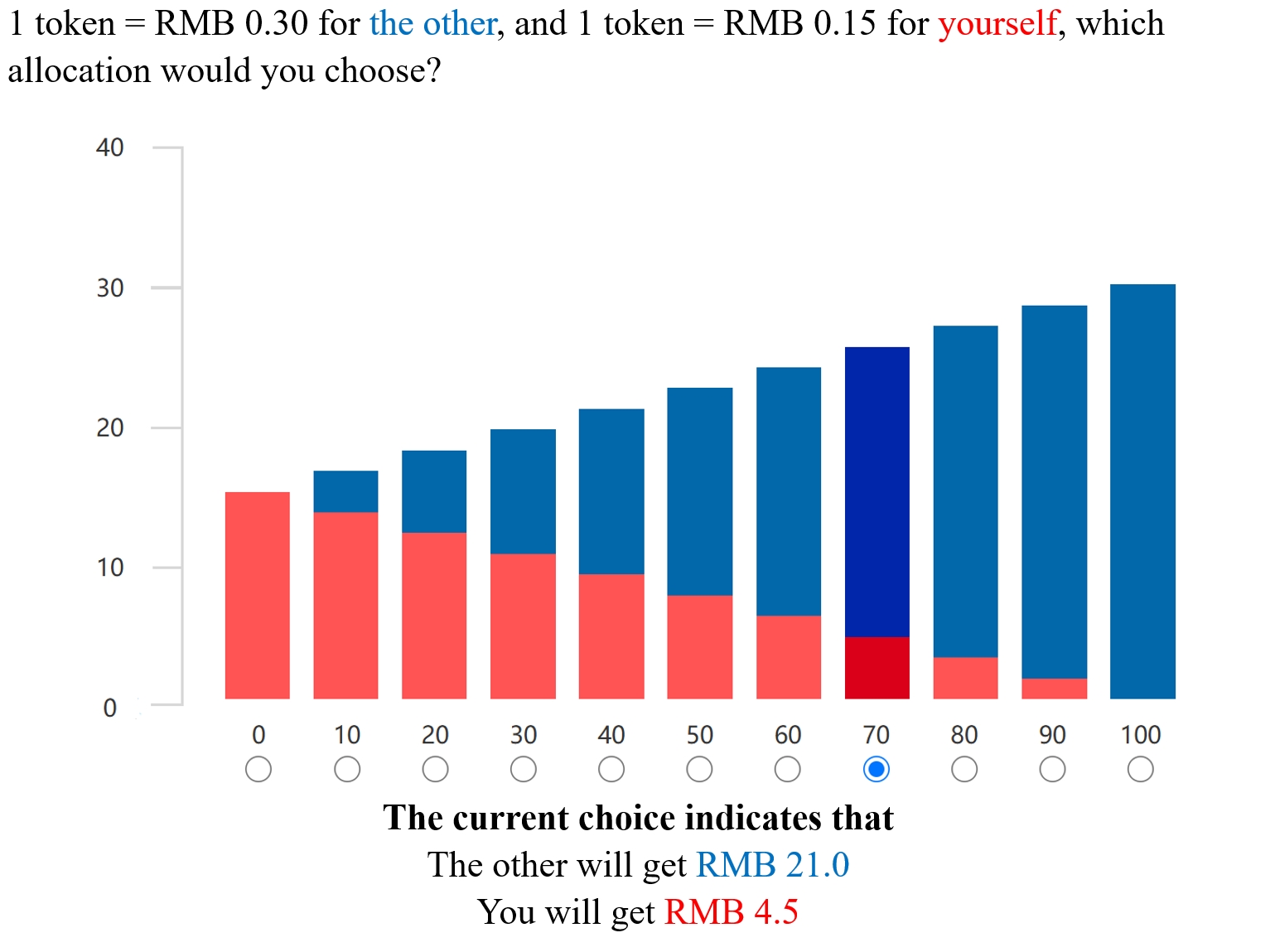}
        \caption{Social Task}
        \label{fig:social}

\end{subfigure}
\begin{subfigure}[b]{0.32\textwidth}

        \includegraphics[width=1\linewidth]{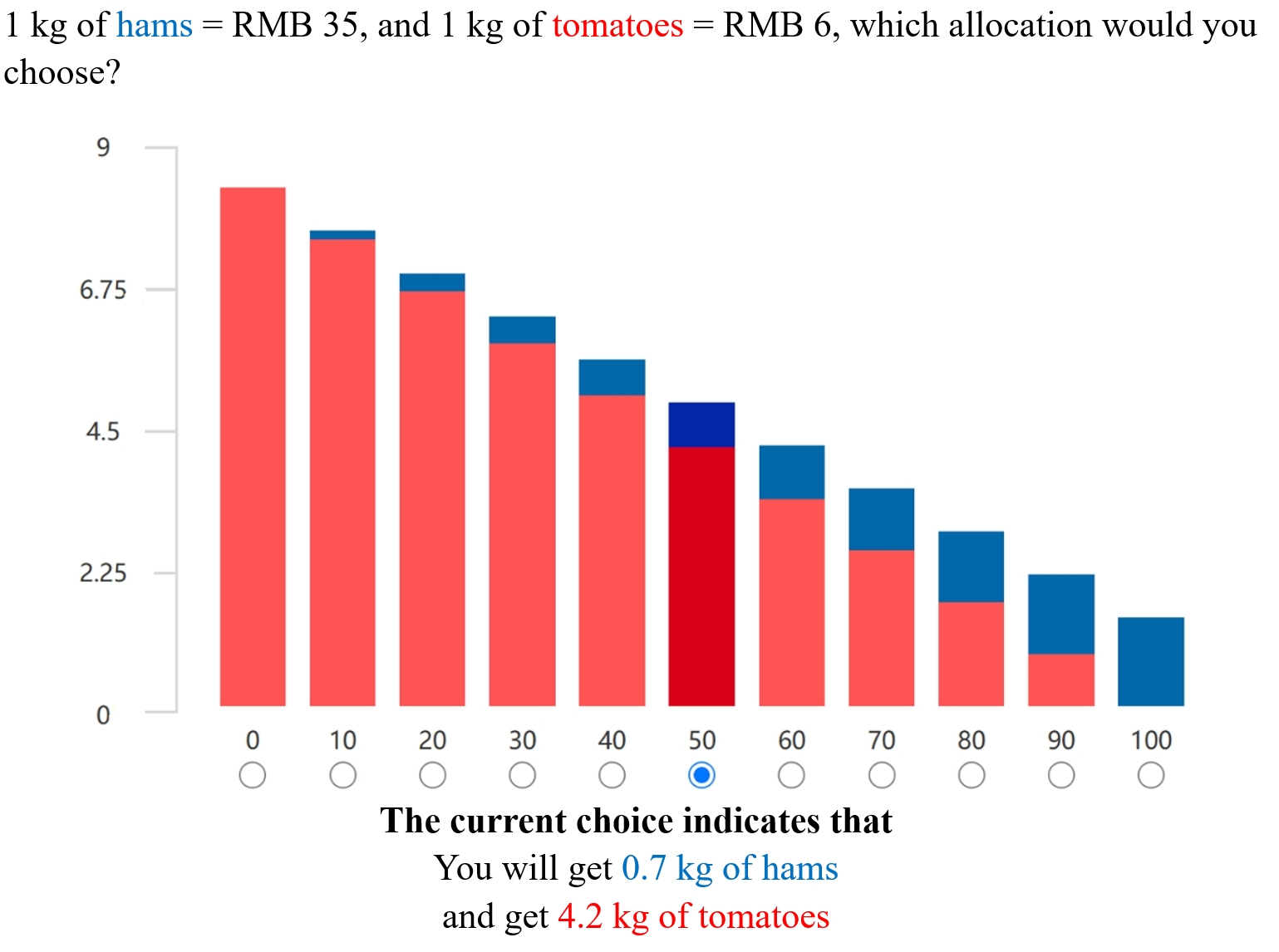}
        \caption{Food Task}
        \label{fig:food}

\end{subfigure}
\caption{Screenshots for the Risk, Social and Food Tasks}
\label{fig:study1_risktask}
\end{figure}

Moreover, to help consumers in the supermarket understand the experimental tasks, we discretize the choice set as in \cite{kim2018role}, by which they choose among 11 allocation options.\footnote{\cite{polisson2013revealed} show that Afriat’s theorem is applicable in discrete settings.} Figure \ref{fig:risk} presents a screenshot of the task. In total, participants face 22 decision tasks presented in random order. Every individual is presented with the same set of budget lines to ensure that the power for revealed preference tests is the same for all participants and to facilitate comparison between participants \citep{halevy2018parametric}. At the end of the experiment, we include a 10-question version of the Big Five personality questions, seven questions from Raven's progressive matrices, and collect demographic information in the post-experiment questionnaire.

To further investigate the correlations of consistency scores between different tasks in the experiment, we conduct Experiment 2. In addition to risky decisions as in Experiment 1, we also measure consistency scores for social and food decisions. We use a modified dictator game to measure consistency scores for social decisions \citep{andreoni2002giving,fisman2007individual}. Participants allocate 100 tokens between themselves and another randomly matched participant in 22 rounds. In each round, each token has a different cash value for the individual and her matched participant. 

To measure consistency scores for food decisions in the experiment, we construct a shopping environment \citep{chen2023emergence,harbaugh2001garp}. In our design, participants allocate the expenditure of RMB 50 (USD 1 $\approx$ RMB 6.5) between a specific type of meat (ham) and a specific type of vegetables (tomato) in 22 rounds. The price level for each product is chosen from the range of actual prices from 2020 to 2021. We present the risk, social, and food tasks in random order to participants and end the experiment with the post-experiment questionnaire identical to that in Experiment 1. Screenshots for the two tasks are presented in Figures~\ref{fig:social} and \ref{fig:food}, and detailed instructions are provided in Appendix~\ref{instructions}. 

The two experiments were conducted from July to September 2021 and in December 2021, respectively. In Experiment~1, of the 6,144 consumers identified as highly frequent consumers introduced in the next subsection, contact information is available for approximately half of them. We invited these consumers to participate in the incentivized experiment in nearby supermarkets. In total, 1,055 highly frequent consumers participated in the experiment in 96 supermarkets. In Experiment 2, another 302 consumers participated in the experiment. To avoid learning spillover between experiments, we only invited consumers who did not participate in Experiment~1. Therefore, most of them are not highly frequent consumers and we do not have enough power to examine the consistency scores of their choices in the supermarket. 

All participants in both experiments received RMB 50 for completing the experiment. Additionally, we randomly paid 10 percent of participants based on their choices. Note that if the decision in the food task was chosen as reward, we deposited redemption vouchers into the consumer's membership account based on her choices. Each individual received an average payment of RMB 52.5 in Experiment 1 and RMB 52.0 in Experiment 2. In Experiment~2, participants also received vouchers redeemable on average for 0.04 kg of hams and 0.10 kg of tomatoes. Table~\ref{tab:demo} reports the summary statistics for our participants in both experiments. For example, among the 1,055 participants in Experiment 1, the average monthly expenditure on meat and vegetables is RMB 482.2, with an average purchase frequency of 11.8 days per month. 

\subsection{Scanner Dataset}\label{sec:data scanner}

We obtained a transaction-level scanner dataset from the company to measure consistency scores in the field. A sample transaction dataset is presented in Table~\ref{tab:sample}, where each observation represents the purchase of a single item, that is, one type of good bought by a consumer in a single shopping instance. The supermarket membership ID uniquely identifies each consumer. The dataset also includes the unique store ID, transaction timestamp, category of the goods, the quantity purchased, the amount of expenditure, and additional transaction details such as discounts.

\begin{table}[hbpt]
\centering
\caption{Sample Records for Scanner Dataset}\label{tab:sample}
\begin{tabular}{cccccc}
\toprule \toprule
\makecell[c]{Membership \\ ID } & \makecell[c]{Store \\ ID } & \makecell[c]{Transaction\\ timestamp } & Category & \makecell[c]{Quantity \\(kg) } & \makecell[c]{Expenditure \\ (RMB)}\\
\midrule
a1 & A1 & 2019-04-13 12:07:10 & Meat & 1.5 & 45.0  \\
a1 & A1 & 2019-09-21 16:35:45 & Meat & 2.0 & 40.0  \\
a1 & A1 & 2019-10-19 10:10:10 & Vegetable & 4.2 & 8.4  \\
a1 & A1 & 2019-11-10 19:05:05 & Vegetable & 1.8 & 3.6  \\
\bottomrule \bottomrule
\end{tabular}
\end{table}

We now introduce how we construct individual-level choice dataset $\{ (p^{t}, x^{t}) \}_{t=1}^{T}$ based on supermarket purchases. We focus on two categories of consumption, meat and vegetables, in the baseline analysis. The choice of these two categories is due to the following two reasons. First, because these two are most frequently purchased goods in the supermarkets, we are able to garner a sufficient number of observations. Second, because the unit for meat and vegetables is kilogram, it is convenient for us to have a refined measure of the quantity. We measure the quantity and price for meat and vegetables by averaging the quantity and prices of items in each category. We denote $q_{in}^t$ as the quantity of item $i$ in consumption category $I$ for consumer $n$ in month $t$, and $Q^t_{In}$ as the quantity of consumption category $I$ for consumer $n$ in month $t$, which is obtained by the sum of $q_{in}^t$. 
$$ Q_{In}^t=\sum_{i\in I} q_{in}^t.$$
In a similar vein, we denote $P_{In}^t$ as the price of the consumption category $I$ for consumer $n$ in month $t$, which is obtained by averaging prices weighted by quantity: 
$$ P_{In}^t=\frac{\sum_{i\in I} e_{in}^t} {Q_{In}^t },$$
where $e_{in}^t$ is the expenditure for item $i$ in category $I$. 
With quantity $Q_{In}^t$ and price $P_{In}^t$, we can construct budget lines for consumer $n$ in month $t$ and perform revealed preference analysis for each consumer $n$.

The construction of budget lines involves certain assumptions and simplifications. First, our approach of forming composite goods and aggregating purchases over time aligns with standard practices in applied demand analysis \citep[e.g.,][]{barten1969maximum,deaton1980almost,allen2024revealed}. Second, utilizing high-frequency individual-level data, we focus on consumers with consecutive purchase records and create an individualized price index for each consumer. This contrasts with studies like \cite{echenique2011money}, which assume the same price for each item for all consumers, and \cite{dean2016measuring}, which assume uniform pricing for all items within a category. Many other studies adopt similar ``same-price" assumptions  \citep[e.g.,][]{allen2020satisficing,demuynck2018revealed,tipoe2021price}. Our individualized price index strengthens the power of GARP tests and reduces measurement errors. Third, we assume preferences for meat and vegetables are weakly separable from other goods and services, a common assumption in applied demand analysis \citep[e.g.,][]{blundell2003nonparametric,dean2016measuring,deaton1980almost,echenique2011money}. This enables independent examination of consumer choices within these categories. In Section~\ref{sec:robustness}, we explore the impact of other goods and income using some revealed preference techniques \citep{brown2007nonparametric, deb2023revealed, cherchye2018normality}. Lastly, we use the final price with discounts to reduce measurement errors, and use shelf prices as a robustness check, with our main results remaining consistent.

Following the above construction of budget lines, we identify 6,144 high-frequency consumers (as noted in Section \ref{sec:risk}) who have purchase records in both categories for 24 consecutive months from January 2019 to December 2020. After matching the scanner data with participants from Experiment~1, we obtain a final dataset of 1,055 consumers, which we refer to as Scanner Dataset~1.

Furthermore, to measure correlations of consistency scores across different categories in the supermarket, we include two additional frequently purchased categories: fruits and snacks, yielding 822 consumers who have purchase records of all four categories for 24 consecutive months in 2019 and 2020.
To measure correlations of consistency scores over time periods, we focus on 938 consumers with purchase records of both meat and vegetables for 48 consecutive months from 2018 to 2021. We refer to this dataset as Scanner Dataset 2.

\section{Results}
\label{sec:results}
\subsection{Experiment 1 vs. Scanner Dataset 1}
\label{sec: lab vs field}

\begin{figure}[htbp]
\centering
\includegraphics[width=0.6\textwidth]{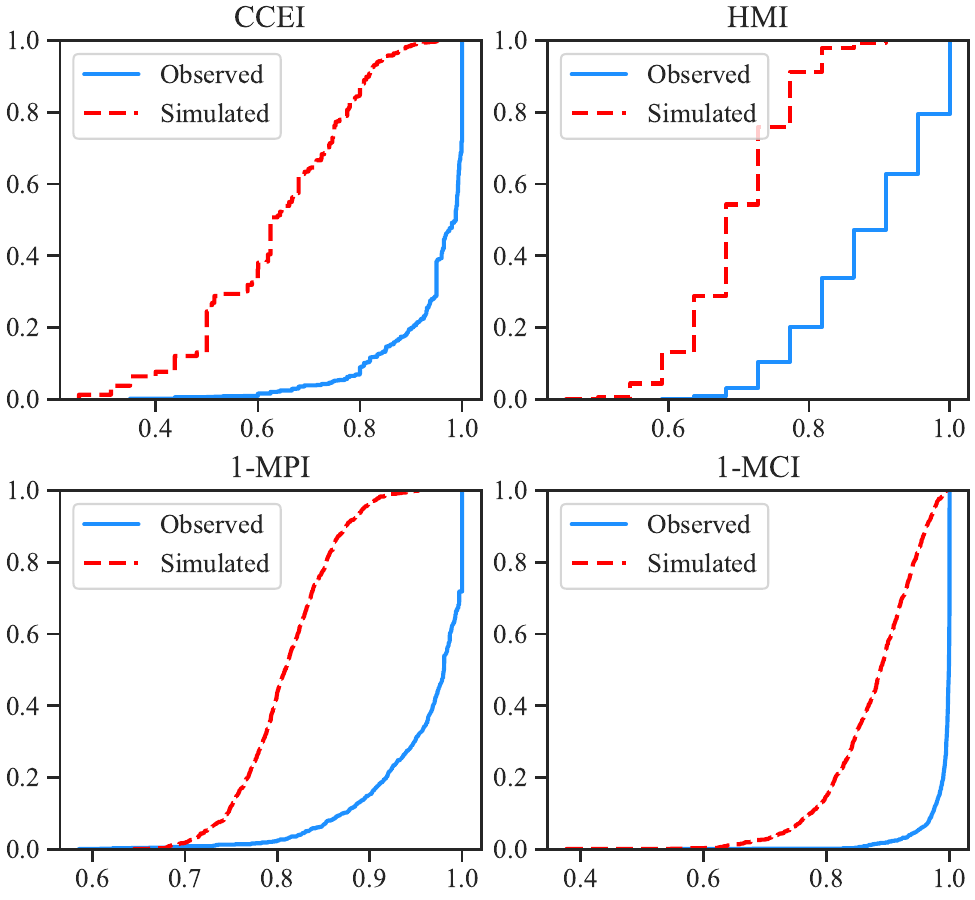}
\caption{Distribution of Consistency Indices for Risky Decisions in Experiment 1}\label{study1_rationality_risk}
\end{figure}

To examine the correlation of choice consistency between experimental and scanner data, we utilize choices made by 1,055 individuals both in Experiment~1 and Scanner Dataset~1. Figure~\ref{study1_rationality_risk} presents the cumulative distribution of consistency indices for risky decisions (solid line) and the summary statistics are in Table~\ref{tab:rationality desc} (Panel A). Taking CCEI as an example, we find that 28.2\% participants do not have GARP violations, and the average CCEI score is 0.941, which implies that the budget must be reduced by 5.9 percent to remove all GARP violations. The observed CCEI distribution is comparable to those in other experimental studies on risky decisions. For example, the average CCEI scores are 0.881 in \cite{choi2014more}, 0.937 in \cite{choi2007consistency}, and 0.979 in \mbox{\cite{halevy2018parametric}}. To examine the power of the test, we calculate simulated scores by generating observations using uniformly random allocations over 11 options in 22 rounds for 1,000 times \citep{bronars1987power}. In contrast to the CCEI scores based on choice data, the data from the simulated random allocations (dashed line in Figure~\ref{study1_rationality_risk}) have substantially lower scores ($p<0.01$, two-sided two-sample $t$-tests). For example, only 1.5 percent of their CCEI scores are greater than 0.9. This suggests that our choice data have sufficient power to detect GARP violations.

\begin{figure}[htbp]
\centering
\includegraphics[width=0.6\textwidth]{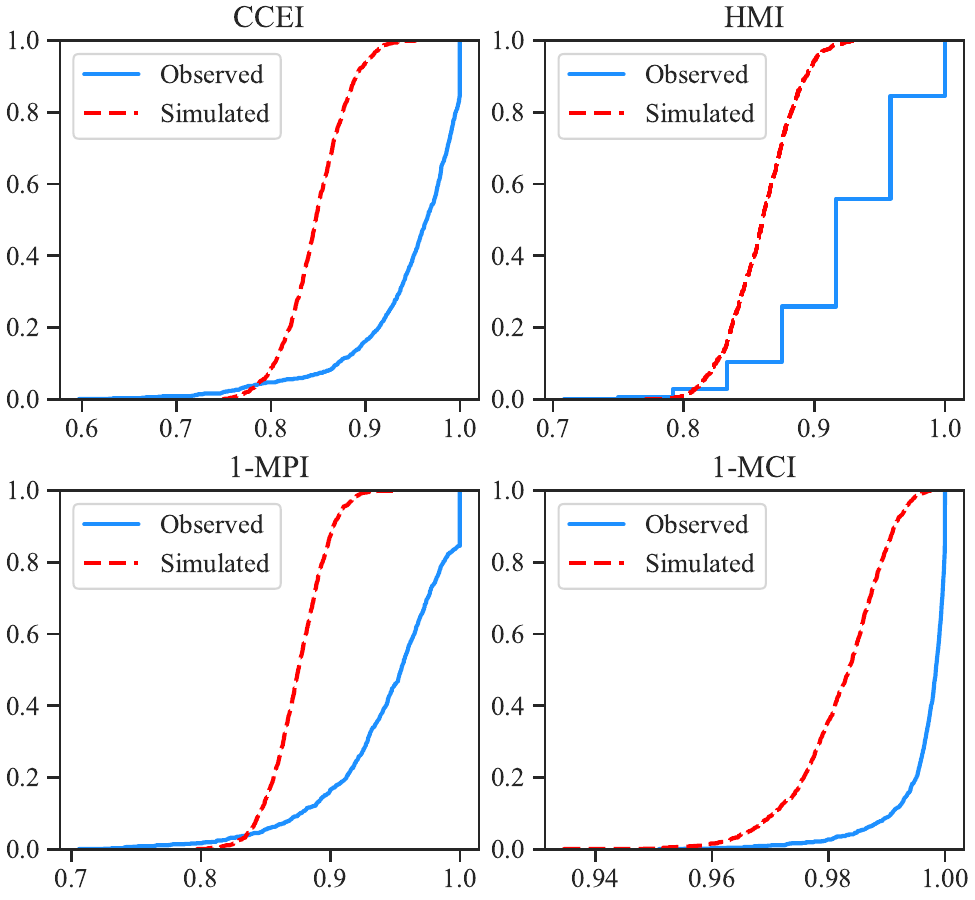}
\captionsetup{justification=centering}
\caption{Distribution of Consistency Indices for Consumption Decisions in Scanner Dataset~1}\label{study1_rationality_food}
\end{figure}

Figure \ref{study1_rationality_food} presents the cumulative distribution of observed (solid line) and simulated (dashed line) scores of four consistency indices for consumption decisions in Scanner Dataset~1 (see Panel B in Table~\ref{tab:rationality desc} for summary statistics). We find that 15.5\% consumers have no GARP violations, and the average CCEI score is 0.946. Using scanner data, the average CCEI scores are 0.976 in \cite{echenique2011money}, and 0.99 in \cite{dean2016measuring}. As supermarket consumers face different budget lines, we calculate the simulated CCEI score for each consumer. Each simulated CCEI score is the average score generated by repeating uniformly random allocations over budget shares along the 24 budget lines of a consumer 100 times \citep{dean2016measuring}. The results show that simulated CCEI scores are significantly smaller than CCEI  scores for actual choices ($p<0.01$, two-sample \textit{t}-tests), and only 6.2 percent of them are over 0.9.\footnote{Using two-sample $t$-tests, all observed scores are significantly higher than simulated scores ($p<0.01$) in both Experiment 2 and Scanner Dataset 2. To avoid redundancy, we omit these repeated expressions in the following analyses.} This confirms the sufficient power of budget lines in the scanner dataset.

We now turn to the main research question regarding the generalizability of consistency measures between risky decisions in the experiment and consumption decisions in the supermarket. We check their association using Spearman's rank correlation ($r$), which is the Pearson correlation between the rank values of those two variables. Moreover, for the correlation coefficients reported in the study, we refer to $[0,0.1)$ as very low or no correlation, and, following the convention proposed by \citet{cohen1988statistical}, we refer to $[0.1,0.3)$ as low, $[0.3,0.5)$ as moderate, and $[0.5,1]$ as high.

\begin{figure}[t]
    \centering
      \includegraphics[width=0.7\textwidth]{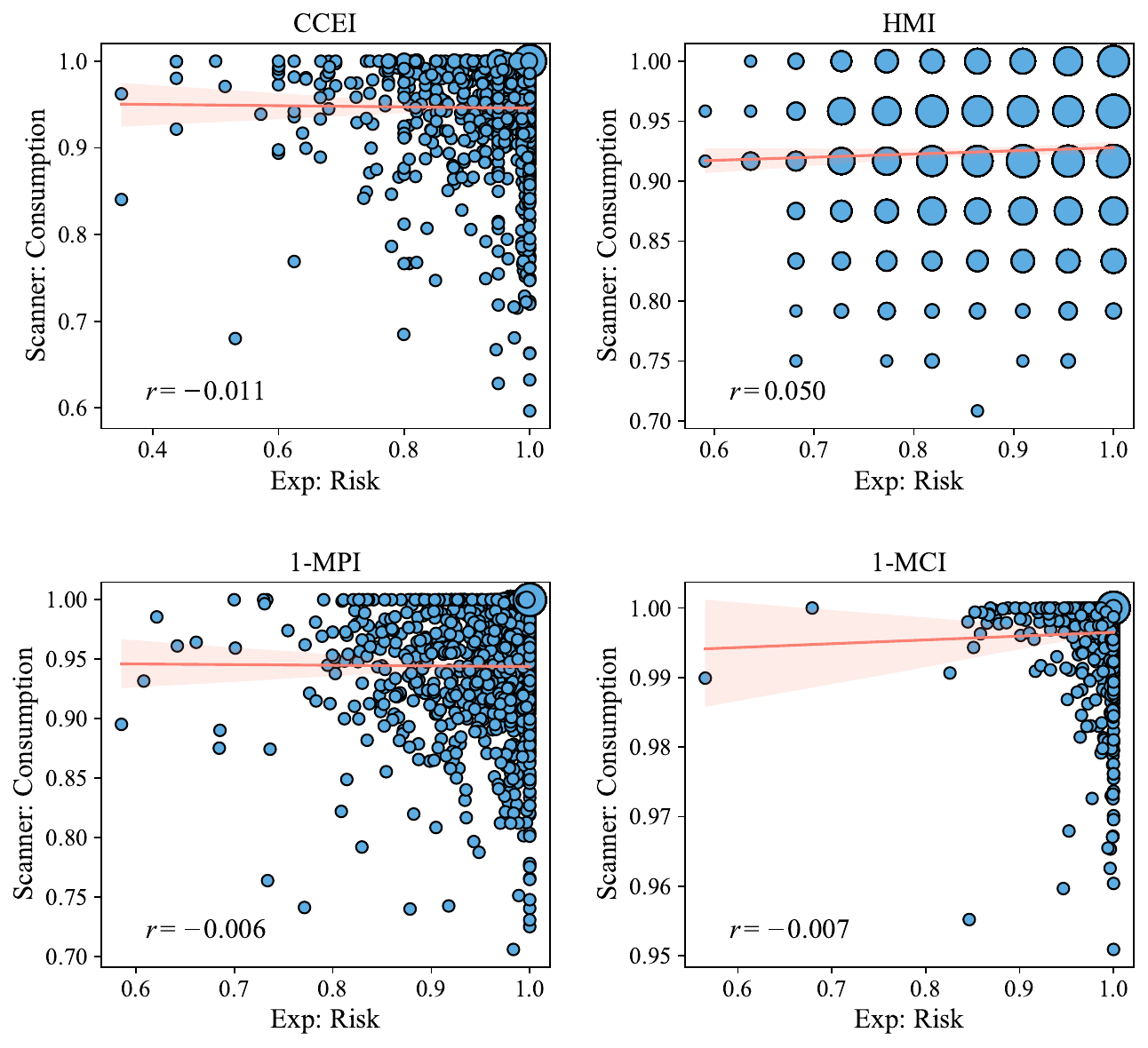}  
    \caption{Correlation of Consistency Indices Between Experiment 1 and Scanner Dataset 1} \label{study1_riskvsfood}
   \begin{minipage}{0.95\textwidth}
        \begin{tablenotes}
\scriptsize \item  \textit{Notes}: $N=1,055$. Spearman’s rank correlations (\textit{r}) are reported.  The size of each circle represents the number of observations at the point. The lines represent the linear fit, and 95\% confidence intervals are shown in the shaded area.  {*}~\(p<0.10\), {**}~\(p<0.05\), {***}~\(p<0.01\).
        \end{tablenotes}
    \end{minipage}
\end{figure}

Figure \ref{study1_riskvsfood} presents scatter plots using four indices, CCEI, HMI, MPI and MCI, and shows that the rank correlation coefficients are very low and statistically insignificant ($p>0.1$ for all indices). That is, the consumer with a high level of consistency score for risky decisions in the experiment does not necessarily exhibit a high level of consistency for consumption decisions in the supermarket. We summarize this finding as follows.

\begin{result}
\label{res:lab vs field}
The consistency scores are not correlated between risky decisions in Experiment 1 and consumption decisions in Scanner Dataset 1.
\end{result} 

Given the above similar correlations among all consistency indices, we examine the relationships among these indices within each setting. Table~\ref{tab:study1_cor1} reports the correlation coefficients for pairwise comparisons and shows that the average of the correlation coefficients is 0.809. Although there have been theoretical discussions on how best to capture the extent of GARP violations \citep{apesteguia2015measure,echenique2021meaning,polisson2024rationalizability}, we show empirically that these indices are highly correlated. In subsequent analyses, for ease of presentation, we focus on results using CCEI and report results for other indices as robustness checks.

\begin{table}[H]
      \begin{center}      
\caption{Correlation of Consistency Indices in Experiment~1 and Scanner Dataset~1} \label{tab:study1_cor1}
    \begin{tabular}{lcccclccc}
\toprule \toprule 
& \multicolumn{3}{c}{Exp: Risk} &     &     & \multicolumn{3}{c}{Scanner: Consumption} \\
\cmidrule{2-4}\cmidrule{7-9}        
& \multicolumn{1}{c}{CCEI} & \multicolumn{1}{c}{HMI} & \multicolumn{1}{c}{1-MPI} &     &     & \multicolumn{1}{c}{CCEI} & \multicolumn{1}{c}{HMI} & \multicolumn{1}{c}{1-MPI} \\
\cmidrule{2-4}\cmidrule{7-9}  
HMI & 0.756*** &  &  &  & HMI & 0.698*** &  &  \\
1-MPI & 0.944*** & 0.644*** &  &  & 1-MPI & 0.890*** & 0.582*** &  \\
1-MCI & 0.988*** & 0.784*** & 0.936*** &  & 1-MCI & 0.905*** & 0.790*** & 0.791*** \\
\bottomrule \bottomrule
    \end{tabular}%
    
    \begin{tablenotes}
\scriptsize
\item \qquad \textit{Notes}: $N=1,055$. Spearman’s rank correlations are reported.  {*} \(p<0.10\), {**} \(p<0.05\), {***} \(p<0.01\). 
      \end{tablenotes}
  \end{center}
\end{table}

\subsection{Experiment 2 \& Scanner Dataset 2}

In this section, we examine the correlations of choice consistency when making budgetary choices in Experiment 2 and consumption choices in Scanner Dataset 2. This analysis helps us see whether the lack of correlation in consistency scores between risky decisions in Experiment 1 and consumption decisions in Scanner Dataset 1 is due to the different types of choices being made or different settings.
 
\begin{figure}[htbp]
    \centering       
    \includegraphics[width=\textwidth]{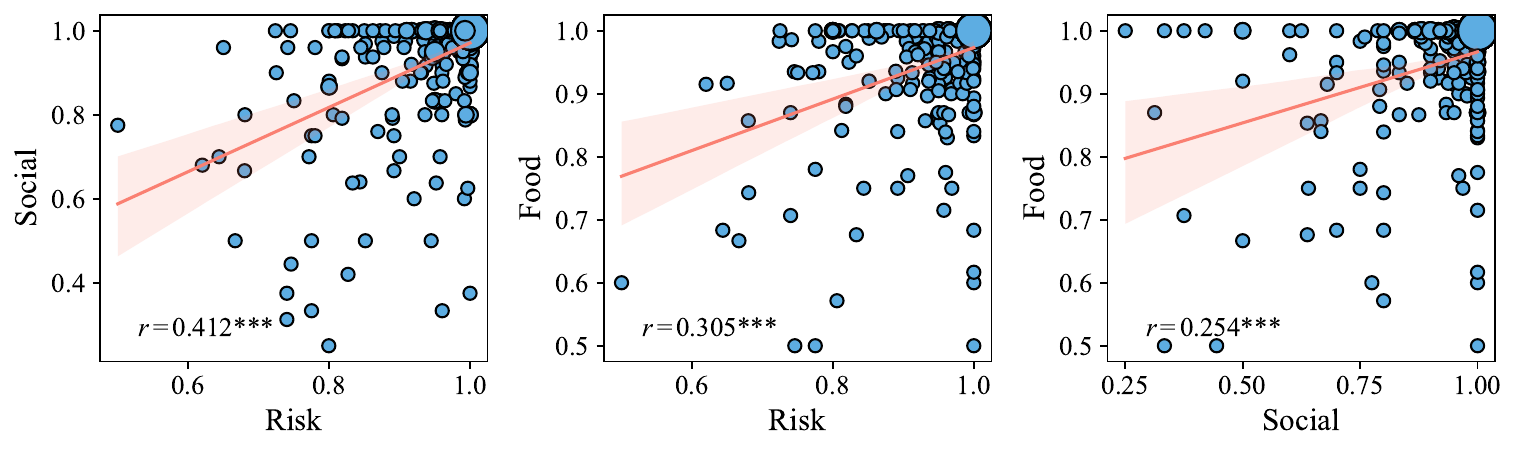}
    \caption{Correlations of CCEI Scores in Experiment 2}\label{study2_correlations}
       \begin{minipage}{0.95\textwidth}
\begin{tablenotes}
    \scriptsize 
      \item \textit{Notes}: $N=302$. Spearman’s rank correlations (\textit{r}) are reported.  The size of each circle represents the number of observations at the point. The lines represent the linear fit, and 95\% confidence intervals are shown in the shaded area. {*}~\(p<0.10\), {**}~\(p<0.05\), {***}~\(p<0.01\).
\end{tablenotes}
    \end{minipage}
\end{figure}

First, we calculate the CCEI scores for each task in Experiment 2 (Figure~\ref{Rationality Index between Domains} and Table~\ref{tab:rationality desc2}) and their patterns are similar to those in Experiment 1. Next, we test pairwise correlations of the CCEI scores between tasks (Figure \ref{study2_correlations}). The results reveal that the CCEI scores for risky decisions are moderately correlated with those for both social decisions (0.412, $p<0.01$) and food decisions (0.305, $p<0.01$) in Experiment~2, suggesting that consumers with higher consistency scores in the risk task tend to exhibit a higher level of consistency in the social and food tasks in the experiment. Meanwhile, the CCEI scores for the social and food decisions are weakly correlated (0.254, $p<0.01$). These lead to our next observation.

\begin{result}\label{res: exp2} The correlations of consistency scores are moderate between risky and social (food) decisions, and low between social and food decisions in the experiment.
\end{result}

\begin{figure}[H]
    \centering
    \includegraphics[width=\textwidth]{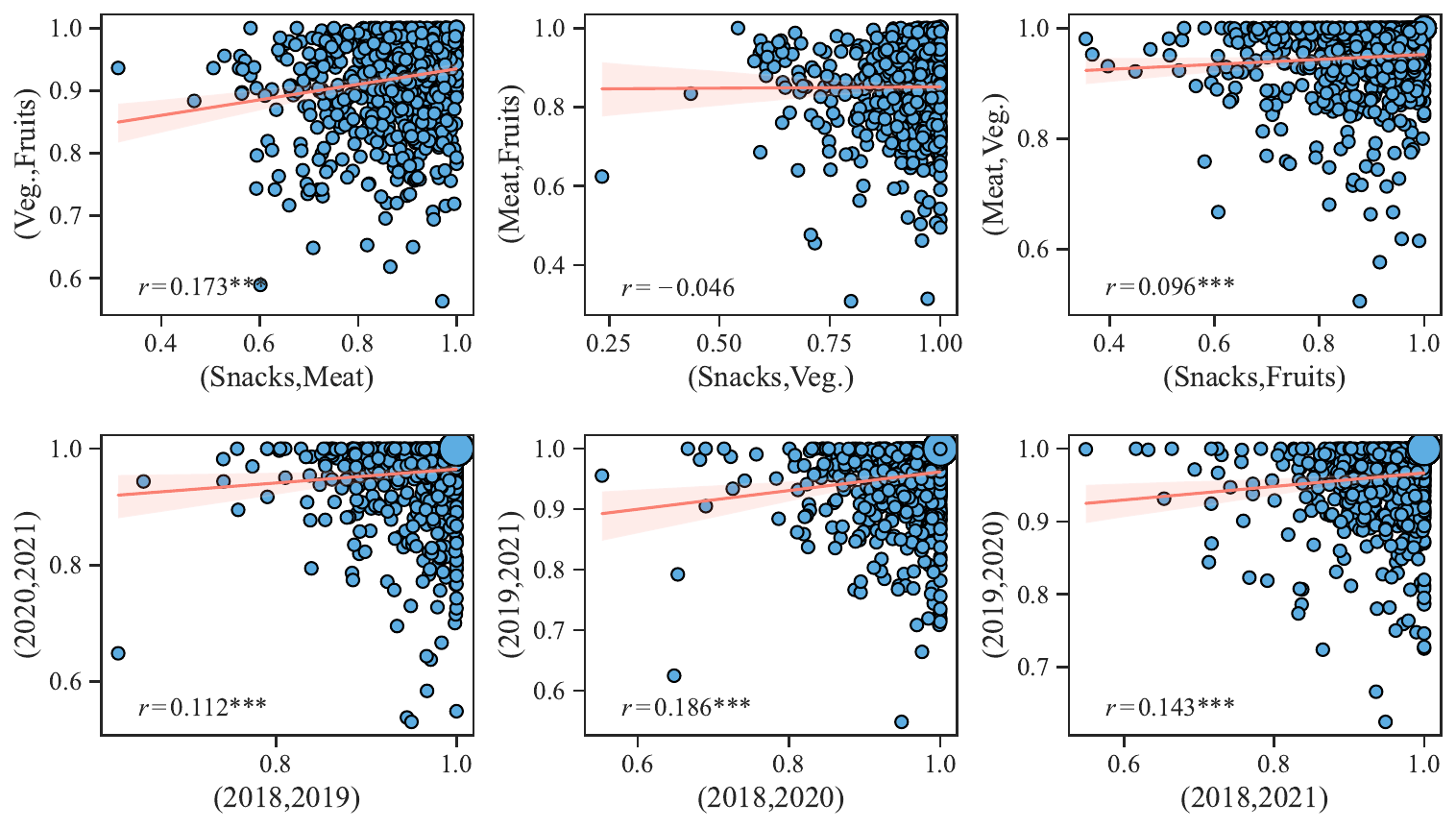}
    \caption{Correlations of CCEI Scores in Scanner Dataset 2}\label{fig:study3_corr}
           \begin{minipage}{0.95\textwidth}
\begin{tablenotes}
    \scriptsize 
      \item \textit{Notes}: {$N=822$ ($938$) for the three correlations across categories (time periods). Spearman’s rank correlations (\textit{r}) are reported.  The size of each circle represents the number of observations at the point. The lines represent the linear fit, and 95\% confidence intervals are shown in the shaded area.  {*}~\(p<0.10\), {**}~\(p<0.05\), {***}~\(p<0.01\).}
\end{tablenotes}
\end{minipage}
\end{figure}

We similarly summarize the consistency of choices in Scanner Dataset 2 (Figure~\ref{fig:study3_des} and Table~\ref{tab:rationality desc3}), and examine correlations of consistency across nonoverlapping pairs of consumption categories and time periods. Note that we do not look at the correlations of consistency scores between pairs that share the same categories (time periods) because the same consumption choices in those overlapping categories (time periods) can create correlations. Figure~\ref{fig:study3_corr} presents Spearman's rank correlations of CCEI scores. The correlations across categories are generally positive and statistically significant (0.173, $p<0.01$; -0.046. $p>0.1$; and 0.096, $p<0.01$). Similarly, the correlations across time windows are consistently positive and significant (0.112, $p<0.01$; 0.186, $p<0.01$; and 0.143, $p<0.01$). We summarize these findings in the following.

\begin{result}
\label{res:scanner2}
The consistency scores show generally positive and significant correlations across consumption categories and time periods in the supermarket, although the magnitude of these correlations is relatively low.
\end{result} 

\subsection{Robustness Checks}
\label{sec:robustness}

In this subsection, we conduct several robustness checks on our main results, including alternative indices of HMI, MPI, and MCI, the power to detect GARP violations, imposing more assumptions on preferences, and fitness in structural estimations.

\textbf{Alternative Indices} In addition to CCEI, alternative indices including HMI, MPI and MCI are proposed to measure the degree of GARP violations. We use these indices to examine the correlations.

\textbf{Heterogeneous Power} 
Consumers face individualized budget sets in our scanner dataset, so the power to detect GARP varies across consumers. To address this issue, we calculate the Selten score by subtracting the simulated CCEI from the observed CCEI \citep{selten1991properties, dean2016measuring}. In addition, we use the residual from a regression of simulated CCEI on observed CCEI (\textit{Power-adjusted CCEI}). 

\textbf{More (Realistic) Assumptions on Preferences} \
Since GARP measures the extent of maximizing some well-behaved, i.e., continuous and strictly increasing, utility function, it is possible that our choice dataset satisfies GARP but may fail to meet requirements on particular preferences. Thus, we consider more realistic restrictions and compute new consistency indices under these conditions. Specifically, two properties related to income are often discussed. First, we examine the adherence to the normality of demand for both goods, suggesting that the demand for commodities increases as income increases, keeping their prices constant (denoted as \textit{Normal}) \citep{cherchye2018normality}.\footnote{\citet{cherchye2018normality} provide the necessary and sufficient condition for the normality of both goods. To further assess how well our data align with this condition, we scale the budget in their NARP-\uppercase\expandafter{\romannumeral 1} condition (Definition~3) by a factor $e$, and identify the largest $e$ that allows the dataset to meet the condition for both goods. We then regard this maximum $e$ as \textit{Normal}.} 
Second, we consider the homothetic preference, assuming a linear relationship between income level and consumption (\textit{Homothetic}) \citep{heufer2019homothetic,varian1983non}. Moreover, consumers' utility may be negatively affected by the expenditure, especially in the supermarket with different expenditures across periods. Therefore, we consider the quasilinear utility $U(x)-px$, where the expenditure of goods reduces the utility in a linear form (\textit{Quasilinear}) \citep{brown2007nonparametric,demetry2022testing}. Meanwhile, to further evaluate choices on two contingent assets in the experiment, we also consider the compliance with first-order stochastic dominance (FOSD)  and expected utility (EU) maximization \citep{nishimura2017comprehensive,polisson2020revealed}.

\textbf{Preference Estimation} \ In addition to the revealed preference analysis, we further measure the extent to which certain specific types of preferences fit the choice data. For risky decisions, we apply the disappointment aversion (DA) model, which is equivalent to rank-dependent utility with binary states. For the other types of decision, we use the utility functions of constant elasticity of substitution (CES) \citep{fisman2007individual}, in which the parameter $\alpha$ captures the relative weight placed on the first good, and $\rho$ is the curvature of indifferent curves. We report the detailed estimation procedure in Appendix~\ref{app:est}. Because the log likelihood (\textit{LL}) measures the goodness of fit of the utility model, we may interpret it as the degree of noise when making decisions.

Table~\ref{tab:summary} summarizes the correlations using these consistency indices (see Tables \ref{tab:mechanism_s1} to \ref{tab:mechanism_s3} for details). We find support for the robustness of Result~\ref{res:lab vs field} on no correlation between experimental and scanner data and Result~\ref{res: exp2} on moderate correlations within experimental data. Notably, after imposing more assumptions on preferences or evaluating the fit of preference structural estimation in the supermarket, correlations within scanner data, as described in Result \ref{res:scanner2}, increase substantially both in magnitude and significance. Therefore, despite the inherent challenges in constructing consumption choices and budget constraints from scanner data, we can recover certain aspects of choice consistency in supermarkets. 

\begin{table}[htbp]
\centering
\caption{Robustness of Correlations of Consistency Indices}
\label{tab:summary}
 \begin{threeparttable}
\begin{tabular}{lc>{\raggedleft\arraybackslash}p{1.5em}>{\centering\arraybackslash}p{1em}p{1.5em}>{\raggedleft\arraybackslash}p{1.5em} >{\centering\arraybackslash}p{1em} p{1.5em}>{\raggedleft\arraybackslash}p{1.5em} >{\centering\arraybackslash}p{1em} p{1.5em}}
\toprule \toprule
 & \multirow{2}{*}{\makecell[c]{Exp. 1 vs. Scanner \\ Dataset 1}} & \multicolumn{3}{c}{\multirow{2}{*}{Exp. 2}} & \multicolumn{6}{c}{Scanner Dataset 2} \\
 \cmidrule{6-11}
 &  & \multicolumn{3}{c}{} & \multicolumn{3}{c}{Across cat.} & \multicolumn{3}{c}{Across time} \\
 \midrule
\multicolumn{11}{l}{\textit{Panel A: Consistency indices}} \\
 \midrule
CCEI & \none & \medium & \medium & \weak & \weak & \none & \none & \weak & \weak & \weak \\
HMI & \none & \medium & \medium & \medium & \none & \none & \none & \none & \weak & \weak \\
MPI & \none & \medium & \medium & \weak & \weak & \none & \none & \weak & \weak & \weak \\
MCI & \none & \medium & \medium & \weak & \weak & \none & \weak & \weak & \weak & \weak \\
 \midrule
\multicolumn{11}{l}{\textit{Panel B: Power of the test}}  \\
 \midrule
Selten score & \none & $-$ & $-$ & $-$ & \none & \none & \none & \none & \weak & \weak \\
Power-adjusted CCEI & \none & $-$ & $-$ & $-$ & \weak & \none & \none & \none & \weak & \weak \\
 \midrule
\multicolumn{11}{l}{\textit{Panel C: More (realistic) assumptions on preferences}} \\
 \midrule
Normal & \none & \medium & \medium & \medium & \weak & \weak & \weak & \medium & \high & \medium \\
Homothetic & \none & \medium & \weak & \weak & \weak & \none & \weak & \weak & \medium & \medium \\
Quasilinear & \none & $-$ & $-$ & $-$ & \weak & \medium & \weak & \weak & \medium & \medium \\
FOSD & \none & \weak & \weak & $-$ & $-$ & $-$ & $-$ & $-$ & $-$ & $-$ \\
EU & \none & \weak & \weak & $-$ & $-$ & $-$ & $-$ & $-$ & $-$ & $-$ \\
 \midrule
\multicolumn{11}{l}{\textit{Panel D: Preference estimation}}  \\
 \midrule
LL & \none & \high & \medium & \medium & \weak & \none & \weak & \medium & \high & \medium \\
\bottomrule \bottomrule
\end{tabular}
		\begin{tablenotes}
			\scriptsize \item \textit{Notes}: \none /\weak /\medium /\high \ represents that the magnitude of correlation coefficient is in [0,0.1)/[0.1,0.3)/[0.3,0.5)/[0.5,1]. Specific values are detailed from Tables \ref{tab:mechanism_s1} to \ref{tab:mechanism_s3}. $-$ denotes that the index is not applicable. Within Experiment~2, the three coefficients represent the correlations of consistency scores between risk and social tasks, risk and food tasks, and social and food tasks, respectively. In Scanner Dataset~2, the three coefficients across categories represent those between (Snacks,Meat) and (Veg.,Fruits), (Snacks,Veg.) and (Meat,Fruits), and (Snacks,Fruits) and (Meat,Veg.); those across time are correlations between (2018,2019) and (2020,2021), (2018,2020) and (2019,2021), and (2018,2021) and (2019,2020).
		\end{tablenotes}
\end{threeparttable}
\end{table}

Additionally, we also conduct a few more robustness checks. First, we use shelf prices instead of final price with discounts. Second, we restrict our analysis to the most frequently purchased subcategory, i.e., pork and leaf vegetables, within each category and calculate the consistency scores accordingly. Third, we assume that the price in each category is the same for all consumers in each month \citep{echenique2011money,dean2016measuring}. Specifically, we obtain the aggregate price index $P_{I}^t$ for category $I$ in month $t$: $P_{I}^t = \frac{\sum_{i \in I} e_{i}^t}{Q_{I}^t}$, where $e_{i}^t$ represents the total expenditure of all consumers on the item $i$ in category $I$ in month $t$, and $Q_{I}^t$ is the total quantity of items purchased in category $I$ in month $t$. In these analyses, all correlation coefficients are close to our baseline findings.

\section{Underlying Mechanisms}
\label{sec:mechanism}

Our findings indicate that, while consistency scores can be derived from choice data across various contexts, they do not always reflect a general ability to make decisions aligned with individual preferences. This resonates with the existing literature that highlights the challenges in inferring true preferences from observable choices. For example, \citet{ariely2003coherent} introduce the notion of coherent arbitrariness, suggesting that subjects are swayed by arbitrary anchors but respond consistently to significant changes in price, quantity, and quality. Similarly, \cite{dean2021credit} argue that intertemporal choices in experiments may not accurately represent true time preferences, as they can be affected by financial shocks and budget constraints. Furthermore, the frequently observed S-shaped probability weighting function in prospect theory may arise from cognitive uncertainty, where individuals tend to default to cognitive shortcuts \citep{enke2023cognitive}, or from the complexities involved in evaluating lotteries \citep{oprea2024decisions}. Observable choices such as those in the Allais paradox and the Ellsberg paradox may also reflect noise or errors \citep{nielsen2022choices,mcgranaghan2024distinguishing} or cognitive challenges such as the failure of contingent thinking \citep{esponda2024contingent,niederle2023cognitive}. Moreover, \cite{declippel2024asif} show that subjects whose choices are consistent with utility maximization often rely on heuristic decision rules that mimic the characteristics of utility functions rather than truly maximize utility. Similarly, the consistency scores derived from observable choices might appear ``as if" they are influenced by unobservable factors such as randomness, budget constraints, preferences, and heuristics, making it challenging to isolate these effects. We will delve into these factors to assess their role in the observed lack of correlation between choice consistency in supermarket and experimental data. This analysis will help shed light on whether the lack of correlation is due to measurement errors or the multidimensional nature of choice consistency across different contexts.

\subsection{Randomness in Choice Behavior}

We examine the randomness of choice behaviors, which can be viewed as a measure of some unobservable factors. We apply a non-parametric one-sided test proposed by \citet{cherchye2023consumers} to examine whether a choice dataset is best explained by a random DM or an approximate utility maximizer. This test uses the permutation approach, which fixes the budget sets but permutes the relative consumption shares over different rounds for each DM, to operationalize the random behaviors. They show that, given a sufficient number of observed choices, the CCEI of an approximate utility maximizer has a probability close to 1 of exceeding the CCEI generated by the permutation test. Consequently, we calculate the proportion of permuted CCEI scores, generated through 10,000 permutations, that surpass the observed CCEI.\footnote{We follow the procedure in \citet{cherchye2023consumers} to abort the test when the proportion exceeds 0.2 for the first 1,000 permutations and report this value to speed up the computation.} 
If such a proportion is smaller than the traditional significance level $\alpha_{\text{sig}}=0.05$, we can regard the DM as an approximate utility maximizer. 

\begin{table}[htbp]
\captionsetup{justification=centering}
  \caption{Proportions of Approximate Utility Maximizers}\label{tab:rejetion rate}
    \centering
    \begin{threeparttable}
\vspace*{-1em}
\begin{tabular}{p{6em}>{\centering\arraybackslash}p{9em}>{\centering\arraybackslash}p{6em}>{\centering\arraybackslash}p{6em}>{\centering\arraybackslash}p{6em}}
 & & & &\\
\toprule \toprule
                     & N          & $\alpha_{\text{sig}} = 0.01$      & $\alpha_{\text{sig}} = 0.05$      & $\alpha_{\text{sig}} = 0.1$     \\
\midrule
\multicolumn{5}{l}{\textit{Panel A: Risk task in Experiment 1 vs. Consumption in Scanner Dataset 1}} \\
\midrule
Risk & 1,055 & 67.87\% & 77.06\% & 82.37\% \\
(Meat,Veg.) & 1,055 & 5.21\% & 22.65\% & 37.82\% \\
\midrule
\multicolumn{5}{l}{\textit{Panel B: Experiment 2}}                                                \\
\midrule
Risk & 302 & 70.86\% & 86.42\% & 89.74\% \\
Social & 302 & 64.57\% & 79.14\% & 83.11\% \\
Food & 302 & 22.52\% & 44.37\% & 55.96\% \\
\midrule
\multicolumn{5}{l}{\textit{Panel C: Scanner Dataset 2}}         \\
\midrule
(Snacks,Meat) & 822 & 4.01\% & 15.94\% & 27.25\% \\
(Snacks,Veg.) & 822 & 4.99\% & 20.44\% & 30.05\% \\
(Snacks,Fruits) & 822 & 3.65\% & 11.80\% & 21.90\% \\
(Meat,Veg.) & 822 & 5.47\% & 20.92\% & 35.16\% \\
(Meat,Fruits) & 822 & 2.80\% & 10.46\% & 18.49\% \\
(Veg.,Fruits) & 822 & 2.31\% & 9.25\% & 16.91\% \\
(2018,2019) & 938 & 8.53\% & 29.64\% & 43.92\% \\
(2018,2020) & 938 & 5.97\% & 24.84\% & 37.31\% \\
(2018,2021) & 938 & 4.58\% & 18.23\% & 31.77\% \\
(2019,2020) & 938 & 5.54\% & 22.60\% & 36.57\% \\
(2019,2021) & 938 & 7.04\% & 24.41\% & 39.13\% \\
(2020,2021) & 938 & 4.90\% & 22.49\% & 35.61\%          \\
\bottomrule\bottomrule
\end{tabular}
\end{threeparttable}
\end{table}

We find that 77.06\% participants in Experiment 1 can be classified as approximate utility maximizers, while this number is 22.65\% when their scanner data are used. Furthermore, an individual classified as an approximate utility maximizer in the experiment does not have a significantly higher chance to be labeled as an approximate utility maximizer in the supermarket (22.5\% vs. 23.1\%, $p>0.1$, two-sample proportion test). We also examine the proportion in Experiment~2 and Scanner Dataset~2 and use alternative significance levels. Table~\ref{tab:rejetion rate} summarizes these proportions with the corresponding significance level and shows that the results are robust.
Our observations are comparable to those of \citet{cherchye2023consumers}, where the proportions are 82\% for the social task in a two-goods setting in the experiment and 30\% for real-life choices at the significance level of 0.05. Overall, this analysis shows that the choices in the scanner data are more likely from random DMs than those in the experiment, and the sources of randomness are distinct in these two settings.

\subsection{Budget Constraints} \label{sec:budget}
An important difference between the two environments is the budget constraint. In the experiment, participants are presented with well-defined budget lines using a simple interface. In contrast, the budget lines of consumers are not directly observable and are constructed by aggregating the choices in a given set of categories. In addition, consumption data often lack the power to detect GARP violations, because budget lines may not cross sufficiently \citep[][Chapter 5]{blundell2003nonparametric,blundell2008best,chambers2016revealed}. However, our scanner data do have the power to detect GARP violations. 
To further take into account perceived budget constraints, 
we utilize two alternative approaches. The first approach is the Price Misperception Index (PMI) in \citet{de2023relaxed}, which quantifies the minimal degree of price misperception that can rationalize choice data.
The second approach is proposed by \citet{deb2023revealed}, which examines the consistency with the Generalized Axiom of Price Preference (GAPP). This approach allows us to calculate deviations from consistent welfare predictions across different budgets, after accounting for one specific type of inattention to prices.\footnote{Given $p^1=(1,2),x^1=(0,2);p^2=(2,1),x^2=(3,1)$, we can find the dataset satisfies GARP. However, it violates GAPP, because the participant can spend less to purchase bundle $x^1$ in period 2, thus being better off at price $p^2$ than at $p^1$. Similarly, she can also spend less to purchase the bundle $x^2$ in period 1 and become better off at $p^1$ than at $p^2$. This leads to a contradiction in the preference for prices. GAPP is a necessary and sufficient condition for the dataset to be rationalized with $U(x,-px):\mathbb{R}_{+}^{K}\times\mathbb{R}_{-}\rightarrow\mathbb{R}$ \citep[Theorem~1,][]{deb2023revealed}. According to Proposition~3 in \citet{deb2023revealed}, since the expenditure in our experiment is fixed, their measure ($\vartheta^*$) is equivalent to CCEI, but this equivalence does not hold in the supermarket setting.} Panel A in Table~\ref{tab:corr_budget_heuristics} shows that the observed correlations are similar to those using CCEI.

\begin{table}[H]
\centering
\caption{Correlations of Alternative Indices for Budget Constraints and Heuristic Rules}
\label{tab:corr_budget_heuristics}
 \begin{threeparttable}
\begin{tabular}{lc>{\raggedleft\arraybackslash}p{1.5em}>{\centering\arraybackslash}p{1em}p{1.5em}>{\raggedleft\arraybackslash}p{1.5em} >{\centering\arraybackslash}p{1em} p{1.5em}>{\raggedleft\arraybackslash}p{1.5em} >{\centering\arraybackslash}p{1em} p{1.5em}}
\toprule \toprule
 & \multirow{2}{*}{\makecell[c]{Exp. 1 vs. Scanner \\ Dataset 1}} & \multicolumn{3}{c}{\multirow{2}{*}{Exp. 2}} & \multicolumn{6}{c}{Scanner Dataset 2} \\
 \cmidrule{6-11}
 &  & \multicolumn{3}{c}{} & \multicolumn{3}{c}{Across cat.} & \multicolumn{3}{c}{Across time} \\
 \midrule
\multicolumn{11}{l}{\textit{Panel A: Budget constraints}} \\
 \midrule
PMI & \none & \medium & \weak & \weak & \weak & \none & \none & \none & \weak & \none \\
GAPP & \none & $-$ & $-$ & $-$ & \none & \none & \none & \weak & \weak & \weak \\
 \midrule
\multicolumn{11}{l}{\textit{Panel B: Heuristic rules}} \\
 \midrule
Downward & \none & \high & \medium & \medium & \none & \none & \none & \none & \weak & \weak \\
\bottomrule \bottomrule
\end{tabular}
		\begin{tablenotes}
			\scriptsize \item \textit{Notes}:  \none /\weak /\medium /\high \ represents that the magnitude of correlation coefficient is in [0,0.1)/[0.1,0.3)/[0.3,0.5)/[0.5,1]. Specific values are detailed in Table~\ref{app:corr_budget_heuristics}. $-$ denotes that the index is not applicable. Within Experiment~2, the three coefficients represent the correlations of consistency scores between risk and social tasks, risk and food tasks, and social and food tasks, respectively. In Scanner Dataset~2, the three coefficients across categories represent those between (Snacks,Meat) and (Veg.,Fruits), (Snacks,Veg.) and (Meat,Fruits), and (Snacks,Fruits) and (Meat,Veg.); those across time are correlations between (2018,2019) and (2020,2021), (2018,2020) and (2019,2021), and (2018,2021) and (2019,2020).
		\end{tablenotes}
\end{threeparttable}
\end{table}

The construction of budget lines is inherently challenging due to various factors that influence both consumer behavior and firm pricing strategies. On the demand side, consumers often respond to price changes in a coarse manner, leading to suboptimal purchasing decisions. For example, they may exhibit left-digit bias, where they overemphasize the significance of changes in the leftmost digit of a price \citep{list2023left}, and tend to evaluate price changes based on their memory of past prices, which can distort their perceptions of current value \citep{bordalo2020memory}. On the supply side, firms frequently engage in strategic pricing and adjust their prices in ways that can manipulate consumer perceptions and purchasing behaviors. As a result, budget constraints are not exogenously given, but influenced by market dynamics. Addressing these complexities requires further research to accurately measure and understand these factors.

\subsection{Formation of Preferences }
The formation of preferences also differs between environments. \cite{plott1996rational} proposes the notion of preference discovery, in which the choice consistency with utility maximization can be understood as a discovery process in which subjects learn their own needs through a reflection and practice process. This perspective highlights the dynamic nature of the formation of preferences, where individuals refine their understanding of their needs and objectives over time. Complementing this view, \cite{cerigioni2021dual}, \cite{deming2021growing} and \cite{ilut2023economic} distinguish between the decision-making ability to perform a routine task and a problem-solving task, and suggests that the formation of preferences may differ depending on the characteristics of the task. In this regard, DM as a consumer is more likely to do a routine task and may be subjective to some preference shocks related to various contextual factors. In contrast, when participants in the experiment face risky decisions and social decisions, they need to learn the abstract experimental rules, as these rules are not part of their daily experience. Therefore, the choice consistency in the experiment may be related to the ability to ``discover'' one’s preference in the new choice environment and ``solve'' the choice problem accordingly. In the following, we present the results that test the effect of some contextual factors.

\paragraph{Preference Shocks in  Scanner Data}
We examine whether consumers' preferences on meat and vegetables change over time (years, seasons, days, and hours) and other contexts (discounted vs. non-discounted purchases) in Scanner Dataset~1. 

To examine the impact of each factor, we first generate the choice datasets, $s_1$ and $s_2$, for two scenarios separately.\footnote{For years, $s_1$ ($s_2$) encompasses 12 choices in 2019 (2020) for each consumer $n$. For seasons, both $s_1$ and $s_2$ represent one season with six choices, yielding six different combinations of $s_1$ and $s_2$ across four seasons. For days, we focus on working days and non-working days. In China, the year 2019 (2020) comprises 250 (251) working days after excluding weekends and holidays. Using purchasing scanner data for working and non-working days separately, we focus on months with positive consumption to reconstruct the monthly quantity $Q_{In}^t$ and price $P_{In}^t$ in each scenario. For hours, we generate $Q_{In}^t$ and $P_{In}^t$ in a similar way using purchases at meal time and non-meal time, respectively. The meal time spans from 10:00 to 14:00 and from 16:00 to 19:00, and the other time is considered non-meal time. Lastly, to examine the impact of promotional discounts, we obtain $s_1$ for discounted purchases and $s_2$ for non-discounted purchases in a similar manner.} 
Then we calculate the consistency scores for each dataset, $\text{CCEI}_{s_1}$ and $\text{CCEI}_{s_2}$, and the combined dataset, CCEI$_{(s_1,s_2)}$. Thus, we can obtain CCEI$_{\text{diff}(s_1,s_2)}$ =  $\min \{\text{CCEI}_{s_1}, \text{CCEI}_{s_2}\} - $ CCEI$_{(s_1,s_2)}$ at the consumer level. By definition, CCEI$_{\text{diff}(s_1,s_2)} \geq 0$. When CCEI$_{\text{diff}(s_1,s_2)} > 0$, there does not exist a well-behaved utility function that can rationalize the two datasets after accounting for the corresponding consistency scores, implying distinct preferences between the two scenarios. However, CCEI$_{\text{diff}(s_1,s_2)} > 0$ may be simply due to the power of the test, namely, the combined dataset has more observations, so it is more likely to fail the test. To account for this, we build a benchmark by randomly allocating the combined dataset into two simulated choice datasets, $s_1'$ and $s_2'$, keeping the same number of choices between $s_1$~($s_2$) and $s_1'$~($s_2'$). Therefore, we obtain $\text{CCEI}_{\text{diff}(s_1',s_2')}$ for each consumer. Repeating this procedure 100 times, we can examine whether the observed CCEI$_{\text{diff}(s_1,s_2)}$ is equal to the average simulated $\text{CCEI}_{\text{diff}(s_1',s_2')}$, denoted as $\overline{\text{CCEI}}_{\text{diff}(s_1',s_2')}$.

Table~\ref{tab:factors_scanner} reports the results. By conducting pairwise comparisons among the four seasons, we observe that consumers' food preferences in spring are notably different from other seasons, especially fall and winter.\footnote{We also apply the method by \citet{echenique2011money}. Specifically, we calculate the proportion of GARP violations across every two choices within a specific pairwise combination, then take the average across consumers. We find that the results are similar (Tables~\ref{tab:season} and \ref{tab:season_compare}). By estimating the CES functions between seasons, we further observe that $\hat{\alpha}$ in spring is significantly lower than in other three seasons ($p<0.01$ for all comparisons, paired $t$-tests), indicating that consumers change their preferences towards vegetables rather than meat in spring (Table~\ref{tab:est_pref season}). There are two possible explanations. First, most vegetables reach their optimal freshness in spring. Second, following Chinese New Year---a significant temporal landmark that marks the end of winter in China, people tend to set a resolution for the new year, such as improving their diet or adopting healthier eating habits.} 
Additionally, our analysis reveals distinct preferences across multiple dimensions: between meal and non-meal time, and between discounted and non-discounted purchases. Meanwhile, we cannot detect significant differences between the year 2019 and 2020 and between working and non-working days. In general, these results suggest that there are contextual factors that change preferences in the supermarket setting. 

\begin{table}[htbp]
\small
    \centering

    \captionsetup{justification=centering}
  \caption{The Impact of Setting-specific Factors on CCEI in Scanner Dataset~1}\label{tab:factors_scanner}
  \begin{threeparttable}
\begin{tabular}{lccccccc}
\toprule \toprule
Category & $s_1$ & CCEI$_{s_1}$ & $s_2$ & CCEI$_{s_2}$ & CCEI$_{\text{diff}(s_1,s_2)} $ & $\overline{\text{CCEI}}_{\text{diff}(s_1',s_2')}   $ & $p$-value \\
\midrule
Season & Spring & 0.995 & Summer & 0.996 & 0.009 & 0.008 & 0.207 \\
Season & Spring & 0.995 & Fall & 0.996 & 0.012 & 0.009 & 0.005 \\
Season & Spring & 0.995 & Winter & 0.995 & 0.012 & 0.010 & 0.004 \\
Season & Summer & 0.996 & Fall & 0.996 & 0.008 & 0.007 & 0.128 \\
Season & Summer & 0.996 & Winter & 0.995 & 0.010 & 0.009 & 0.102 \\
Season & Fall & 0.996 & Winter & 0.995 & 0.009 & 0.008 & 0.971 \\
Year & 2019 & 0.978 & 2020 & 0.986 & 0.021 & 0.019 & 0.105 \\
Working day & Yes & 0.935 & No & 0.929 & 0.033 & 0.031 & 0.207 \\
Meal time & Yes & 0.937 & No & 0.935 & 0.035 & 0.030 & 0.004 \\
Discount & Yes & 0.932 & No & 0.932 & 0.033 & 0.030 & 0.028 \\
\bottomrule   \bottomrule
    \end{tabular}%
    \begin{tablenotes}
\scriptsize
\item \textit{Notes}: \textit{p}-values of paired \textit{t}-tests are reported.
      \end{tablenotes}
    \end{threeparttable}
\end{table}%

\paragraph{Learning in Experiments}
We analyze whether there is a learning effect in the experiment. We split 22 rounds of choices in Experiment~1 into the first half and the second half, and find that CCEI scores are lower for the first half of risky decisions (0.967 vs. 0.981, $p<0.01$, paired $t$-test).\footnote{While the difference appears to be modest, it is economically meaningful, namely, a 42.42\% reduction in the waste of budget. We also compute the CCEI in \citet{choi2007consistency} and \citet{choi2014more} and find the same result that the CCEI scores for the first half are significantly lower than those for the second half (\citet{choi2007consistency}: 0.960 vs. 0.986, $p<0.01$; \citet{choi2014more}: 0.942 vs. 0.949, $p=0.025$, paired $t$-tests).} Using the same method as mentioned above to calculate and compare CCEI$_{\text{diff}(s_1,s_2)} $ and $\overline{\text{CCEI}}_{\text{diff}(s_1',s_2')}$, the underlying preferences for the two halves are also different, contributing to the deviation from utility maximization. Furthermore, given that we randomize the order of the three tasks in Experiment~2, we find that the Spearman rank correlation between the first and second tasks, and that between the first and third are both 0.284 ($p<0.01$), while that between the last two tasks is 0.350 ($p<0.01$). These findings suggest that generalizability increases as participants gain more experience throughout the experiment.

\paragraph{Individual Characteristics} 
We further examine how individual characteristics affect consistency scores in these two settings in Table~\ref{tab:ols_behavioral}. Independent variables include Raven's IQ test score, shopping frequency, Big Five personality traits, as well as demographic variables that include gender, family income, education, age, and family size. The dependent variables are CCEI scores (Column~1) and consistency with FOSD (Column~2) in Experiment~1, as well as the CCEI scores in Scanner Dataset~1 (Column 3). 

Our results show that the estimated coefficient of IQ is significantly positive for the consistency with FOSD, which is similar to the observation from \citet{cappelen2023development}. In addition, consumers with more shopping experience have significantly higher CCEI scores in the supermarket. This is in line with \cite{list2003does}, showing that individuals with more trading experiences exhibit less endowment effect and behave more closely to neoclassical predictions, and suggesting that consumers can learn to overcome the endowment effect. Given that the maximum score for these measures is 1 and their distributions are skewed, we also apply a Tobit model and use relative rankings as dependent variables separately and show that the results remain robust (Tables~\ref{tab:tobit_behavioral} and \ref{tab:rank_behavioral}).

Additionally, Column~4 builds on Column~3 by adding the CCEI score in Experiment~1 as an independent variable. The coefficient on this variable is statistically insignificant. According to the Frisch-Waugh-Lovell theorem \citep{frisch1933partial,lovell1963seasonal}, this regression result is equivalent to showing that the residuals from regressing CCEI in Experiment~1 on individual characteristics are not significantly correlated with the residuals from regressing CCEI in Scanner Dataset~1 on the same set of characteristics. That is, Result~\ref{res:lab vs field} continues to hold after accounting for individual characteristics in each setting.

\begin{table}[htbp]
\captionsetup{justification=centering}
  \caption{OLS Regressions for Consistency Indices in Experiment~1 and Scanner Dataset~1}
  \label{tab:ols_behavioral}
    \centering
    \sizea
\begin{threeparttable}
\begin{tabular}{p{9em}>{\centering\arraybackslash}p{7em}>{\centering\arraybackslash}p{7em}>{\centering\arraybackslash}p{7em}>{\centering\arraybackslash}p{7em}} 
\toprule\toprule
 & \makecell[c]{Exp:\\ CCEI} & \makecell[c]{Exp:\\ FOSD} & \makecell[c]{Scanner:\\ CCEI} & \makecell[c]{Scanner:\\ CCEI} \\
 & (1) & (2) & (3) & (4) \\
\midrule
\textit{IQ} & 0.003 & 0.013*** & -0.000 & -0.000 \\
\textit{} & (0.002) & (0.004) & (0.001) & (0.001) \\
\textit{Frequency} & 0.059 & 0.057 & 0.241*** & 0.241*** \\
\textit{} & (0.074) & (0.155) & (0.042) & (0.042) \\
\textit{Conscientiousness} & 0.002 & 0.003 & -0.002 & -0.002 \\
\textit{} & (0.002) & (0.003) & (0.001) & (0.001) \\
\textit{Extraversion} & 0.000 & -0.001 & 0.001 & 0.001 \\
\textit{} & (0.002) & (0.003) & (0.001) & (0.001) \\
\textit{Agreeableness} & -0.003 & -0.004 & 0.000 & 0.000 \\
\textit{} & (0.002) & (0.004) & (0.001) & (0.001) \\
\textit{Openness} & -0.003* & -0.005 & 0.002* & 0.002* \\
\textit{} & (0.002) & (0.004) & (0.001) & (0.001) \\
\textit{Neuroticism} & -0.004*** & -0.006* & -0.000 & -0.000 \\
\textit{} & (0.002) & (0.004) & (0.001) & (0.001) \\
\textit{Female} & 0.000 & 0.027 & -0.009* & -0.009* \\
\textit{} & (0.007) & (0.017) & (0.005) & (0.005) \\
\textit{Medium Income} & -0.004 & 0.013 & -0.002 & -0.002 \\
\textit{} & (0.008) & (0.017) & (0.005) & (0.005) \\
\textit{High income} & 0.012 & 0.043** & -0.003 & -0.003 \\
\textit{} & (0.009) & (0.018) & (0.005) & (0.005) \\
\textit{Medium education} & -0.004 & -0.004 & -0.003 & -0.003 \\
\textit{} & (0.008) & (0.016) & (0.004) & (0.004) \\
\textit{High education} & 0.004 & 0.022 & -0.005 & -0.005 \\
\textit{} & (0.008) & (0.017) & (0.006) & (0.006) \\
\textit{Medium age} & 0.016 & 0.055* & -0.004 & -0.004 \\
\textit{} & (0.016) & (0.033) & (0.008) & (0.008) \\
\textit{Elder age} & 0.014 & 0.062* & -0.005 & -0.005 \\
\textit{} & (0.017) & (0.035) & (0.009) & (0.009) \\
\textit{Family size} & -0.002 & -0.004 & 0.001 & 0.001 \\
\textit{} & (0.003) & (0.007) & (0.002) & (0.002) \\
\textit{Simulated CCEI} &  &  & 0.406*** & 0.406*** \\
 &  &  & (0.052) & (0.052) \\
\textit{Exp: CCEI} &  &  &  & -0.005 \\
\multicolumn{1}{c}{} &  &  &  & (0.023) \\
\textit{Constant} & 0.910*** & 0.673*** & 0.587*** & 0.592*** \\
 & (0.025) & (0.054) & (0.047) & (0.051) \\
\midrule
N & 1,055 & 1,055 & 1,055 & 1,055 \\
\textit{R$^2$} & 0.023 & 0.034 & 0.084 & 0.084        \\
 \bottomrule\bottomrule
    \end{tabular}%
    \begin{tablenotes}
    \scriptsize \item \textit{Notes}: 
    \textit{IQ} is the number of correct answers in a seven-question version of Raven’s Progressive Matrices. \textit{Frequency} is the average monthly shopping days (divided by 100) for meat and vegetables from 2019 to 2020. Dummy for medium (high) income indicates family income between RMB 5,001 and 10,000 (more than RMB 10,000) per month. Dummy for medium (high) education indicates education degree of high school (above high school). Dummy for medium (elder) age refers to those born between 1970 and 1989 (before 1969). In the scanner dataset, simulated CCEI is added as an additional control to address the heterogeneous power issue. Robust standard errors are in parentheses. {*}~\(p<0.10\), {**}~\(p<0.05\), {***}~\(p<0.01\).
      \end{tablenotes}
    \end{threeparttable}
\end{table}

\subsection{Heuristic Rules} 

In addition to preferences, decision makers may use different heuristic rules in response to different environments. In the environment of budgetary choices, \cite{choi2006substantive} show that the portfolio choices of subjects can be explained by some simple rules, such as a diversification heuristic, i.e., allocating the same number of points to the two accounts. \cite{halevy2024identifying} design simple investment rules to select portfolios and show that most subjects prefer to make allocations through the rule-based interface. It is possible that in the abstract environment of experiments, participants find the decisions to be difficult and adopt a set of heuristic rules to simplify their choices. In food purchase decisions, consumers may use different heuristic rules, such as habits formed in their daily life \citep{havranek2017habit}, or reference dependence to perceive price differences \citep{thaler1981economic}. 

To explore such heuristic rules, in Experiment~2, we focus on ``middle choosers'' who choose near-middle options when the prices for two goods are similar.\footnote{Specifically, we define middle choosers as participants who select the near-middle options---allocating between 40\% to 60\% budget to both goods (choosing the 5th, 6th or 7th option on the interface)---across six rounds in the risk task or three rounds in the social task where price ratios ($p_1/p_2$) range from 0.9 to 1.1. We cannot identify middle choosers in the food task as no price ratios fall within the range.} The results reveal that middle choosers constitute 26.49\% of participants in both risk and social tasks. Notably, conditional on being middle choosers in the risk task, the likelihood of being middle choosers in the social task is 45.00\%, while this likelihood is only 19.82\% among nonmiddle choosers in the risk task ($p<0.01$, two-sample proportion test).

Another possible heuristic is ``downward-sloping demand", considering the degree of demand change in response to price change. Specifically, we calculate the Spearman's rank correlation between $\ln(x_1/x_2)$ and $\ln(p_1/p_2)$ for each individual to measure compliance with ``downward-sloping demand'' property \citep{echenique2023approximate}. The results show that pairwise correlations range from 0.324 to 0.539 in Experiment~2 ($p<0.01$ for all combinations) (See Panel B in Table~\ref{tab:corr_budget_heuristics}).\footnote{In addition, we also observe certain correlations of estimated parameters, $\hat{\alpha}$ and $\hat{\rho}$, across tasks within Experiment~2 (Table~\ref{tab: corr_para}).} This is notable because individuals do not necessarily respond consistently to prices due to specific utility functions in particular preference tasks. In contrast, ``downward-sloping demand" measures have low or very low correlations in the scanner data. 

As the selection of middle options and the adherence to downward-sloping demand can be rationalized by certain utility functions, the observed correlations may be influenced by an underlying connection among risk, social, and food preferences. However, we believe that these correlations are more likely to arise from similar decision-making modes or heuristics used in response to the comparable budgetary interface presented in risk, social, and food tasks.

\subsection{Summary with a Conceptual Framework}
To help understand our findings, we consider a conceptual framework as follows. The measured choice consistency $e$ is determined by the true unobserved choice consistency $e^*$, the preference factor $\theta$, and the measurement error $\nu$, which is due to measurement. Accordingly, we have 
\begin{equation}
    e_{ij}=f(e^*_{ij},\theta_{ij},\nu_{ij}),
\end{equation}
where $i=1$ refers to the experimental setting, $i=2$ refers to the supermarket setting, and $j$ denotes the preference domain.

Measurement error is likely to be small in the experimental environment as evidenced by the small number of random DMs shown in Table~\ref{tab:rejetion rate}. We observe in Result~\ref{res: exp2} that there are moderate correlations of observed choice consistency between different preference domains. Since risk, social and food preferences are presumably independent, the observed correlations between $e_{1j}$ are more likely to be driven by the underlying correlations of $e^*_{1j}$ between domains. That is, there is a common underlying choice consistency across preferences measured in the experimental setting.

Conversely, in the supermarket context, measurement error is likely to be substantial, given the concerns regarding the construction of budget lines discussed in Section~\ref{sec:budget} and the setting-specific factors outlined in Table~\ref{tab:factors_scanner}. Nevertheless, the correlations between category combinations remain significant, albeit small (Result~\ref{res:scanner2}), and increase to 0.5 when more realistic preference restrictions are applied (Tables~\ref{tab:summary} and \ref{tab:mechanism_s3}). These observations indicate that while measurement error is a non-negligible issue in the scanner data, it is not large enough to completely obscure the correlations of actual choice consistency. Therefore, measurement error alone cannot fully account for the lack of correlation between the experimental and scanner data. Collectively, these findings suggest that the underlying choice consistency is likely uncorrelated between settings, highlighting the multidimensional nature of choice consistency across different domains.

\section{Predictive Validity}
\label{sec:validity}
Our observations are in line with the notion that decision making capacity is multidimensional. To further substantiate our interpretation and to rule out measurement error as a full explanation, we examine the predictive validity of choice consistency in the two settings. Namely, we go beyond their specific measurement domains, and test whether these measures independently predict different aspects of consumer behavior. As \cite{yariv2025intro} note, ``a measure has high predictive validity if it correlates with other variables that are known or expected to be theoretically related to it''. The degree of predictive validity can help shed light on whether the measure reflects meaningful variations in behavior rather than random or idiosyncratic noise. 

In budgetary experiments, choice consistency reflects individuals’ capacity to solve novel problems and identify relevant trade-offs, and it is theoretically and empirically linked to financial consequences such as wealth accumulation \citep{choi2014more,carvalho2024misfortune}. We therefore investigate whether consistency in the experiment predicts money-saving behavior. Money-saving behavior is measured by the frequency and magnitude of discount usage, including proportion of transactions with discounts, aggregate discount rate, and average transaction-level discount rate. 

In the daily supermarket shopping, by contrast, choice consistency indicates the ability to form habits and to maintain routines across time periods in familiar environments \citep{cherchye2020revealed,crawford2010habits}. Consequently, we examine how consistency in the supermarket relates to consumption regularity. Consumption regularity is measured by month-to-month volatility of shopping patterns across hours of day, days of week and ten-day periods of the month. In particular, for each individual in each year, let $\pi_{g}^{t}= \frac{v_{g}^{t}}{\sum_{g\in G} v_{g}^{t}}$, where $g$ denotes the group in the set $G$, $t$ denotes the month, and $v$ represents transaction amount or count. We consider three sets of groups separately defined along the temporal dimensions: hours of day, days of week, and ten-day periods of the month. Within group set $G$, we first calculate the standard deviation of each group $g$'s shares over 12 months, $s_{g} = \operatorname{sd} (\left\{\pi_{g}^{t}  \right\}_{t=1}^{12}  )$, then average these standard deviations across all groups: $V_{G} = \frac{1}{|G|} \sum_{g\in G} s_{g}$. Higher $V_G$ indicates greater instability. 

We link the consistency scores from Experiment~1 and Scanner Dataset~1 to consumers’ supermarket scanner records covering all product categories both within-sample (2019–2020) and out-of-sample (2021). We consider consumers’ in‑store purchases across all product categories. To mitigate the influence of outliers, we further exclude the top 0.5\% transactions according to the transaction amount prior to discounts. In all regression analyses, we control for demographics, IQ, and Big Five personality traits.

We find a clear dissociation in Table~\ref{tab:predictive_validity}. First, consistency in the experiment predicts money-saving behavior (Panel~A). Namely, individuals with higher consistency scores with FOSD are significantly more likely to use discounts.\footnote{Statistically significant effects for experimental measures are found with stricter criteria such as consistency with FOSD or EU maximization; the GARP alone shows no significant effect, which supports our mechanisms above: some individuals may use simple heuristics that satisfy GARP in the experiments but do not necessarily indicate high quality.} This result is economically meaningful, robust across multiple discount metrics, and holds both within-sample (2019–2020) and out-of-sample (2021). In particular, a one-standard-deviation increase in FOSD is associated with a 7.1–7.8\% standard-deviation increase in the proportion of transactions with discounts, a 3.6–7.2\% standard deviation increase in aggregate discount rates, and a 4.9–7.4\% standard-deviation increase in average transaction-level discount rate. In  contrast, supermarket-based consistency measures rarely show a relationship with discount usage.

\begin{table}[H]
\captionsetup{justification=centering}
  \caption{OLS Regressions for Money-saving Behavior and Consumption Regularity}
  \label{tab:predictive_validity}
    \centering 
    
     \sizea
     \makebox[\textwidth][c]{
\begin{threeparttable}
\setlength{\tabcolsep}{2.5pt}
  \begin{tabular}{lccccccccc}
 \toprule \toprule
\multicolumn{10}{l}{\textit{Panel A: Money-saving behavior}} \\
\midrule
 & \multicolumn{3}{c}{\makecell[c]{Proportion of transactions \\ with discounts}} & \multicolumn{3}{c}{Aggregate discount rate} & \multicolumn{3}{c}{\makecell[c]{Average transaction-level \\ discount rate}} \\
\cmidrule(lr){2-4}\cmidrule(lr){5-7}\cmidrule(lr){8-10}
& 2019 & 2020 & 2021 & 2019 & 2020 & 2021 & 2019 & 2020 & 2021 \\
 & (1) & (2) & (3) & (4) & (5) & (6) & (7) & (8) & (9) \\
\midrule
\textit{Exp: FOSD} & 0.045** & 0.043*** & 0.053*** & 0.015 & 0.027** & 0.020* & 0.019* & 0.021** & 0.029*** \\
\textit{} & (0.019) & (0.017) & (0.020) & (0.012) & (0.011) & (0.011) & (0.011) & (0.011) & (0.011) \\
\textit{Scanner: CCEI} & -0.062 & -0.045 & -0.046 & -0.072* & -0.049 & 0.020 & -0.048 & -0.026 & -0.040 \\
\textit{} & (0.066) & (0.059) & (0.069) & (0.039) & (0.035) & (0.036) & (0.037) & (0.035) & (0.040) \\
\textit{Constant} & 0.462*** & 0.435*** & 0.439*** & 0.202*** & 0.163*** & 0.115*** & 0.123*** & 0.106*** & 0.127*** \\
 & (0.072) & (0.064) & (0.076) & (0.044) & (0.038) & (0.041) & (0.041) & (0.038) & (0.044) \\
\midrule
Controls & Y & Y & Y & Y & Y & Y & Y & Y & Y \\
N & 1,052 & 1,046 & 1,038 & 1,052 & 1,046 & 1,038 & 1,052 & 1,046 & 1,038 \\
\textit{R$^2$} & 0.101 & 0.069 & 0.051 & 0.112 & 0.064 & 0.041 & 0.121 & 0.068 & 0.054 \\
\midrule
\multicolumn{10}{l}{\textit{Panel B: Consumption regularity}} \\
\midrule
 & \multicolumn{3}{c}{Hours of day} & \multicolumn{3}{c}{Days of week} & \multicolumn{3}{c}{Ten-day periods of the month} \\
\cmidrule(lr){2-4}\cmidrule(lr){5-7}\cmidrule(lr){8-10}
& 2019 & 2020 & 2021 & 2019 & 2020 & 2021 & 2019 & 2020 & 2021 \\
 & (1) & (2) & (3) & (4) & (5) & (6) & (7) & (8) & (9) \\
\midrule
\textit{Exp: FOSD} & 0.003 & 0.001 & 0.000 & 0.008 & -0.001 & -0.013 & 0.005 & 0.018 & -0.000 \\
\textit{} & (0.003) & (0.004) & (0.004) & (0.007) & (0.009) & (0.009) & (0.010) & (0.015) & (0.017) \\
\textit{Scanner: CCEI} & -0.043*** & -0.039*** & -0.035*** & -0.072*** & -0.075** & -0.077*** & -0.124*** & -0.116** & -0.101** \\
\textit{} & (0.010) & (0.012) & (0.012) & (0.023) & (0.029) & (0.027) & (0.036) & (0.049) & (0.050) \\
\textit{Constant} & 0.115*** & 0.118*** & 0.120*** & 0.212*** & 0.241*** & 0.265*** & 0.329*** & 0.333*** & 0.385*** \\
 & (0.012) & (0.013) & (0.014) & (0.025) & (0.032) & (0.030) & (0.040) & (0.054) & (0.054) \\
\midrule
Controls & Y & Y & Y & Y & Y & Y & Y & Y & Y \\
N & 1,050 & 1,043 & 1,017 & 1,050 & 1,043 & 1,017 & 1,050 & 1,043 & 1,017 \\
\textit{R$^2$} & 0.072 & 0.081 & 0.058 & 0.049 & 0.049 & 0.049 & 0.063 & 0.044 & 0.047        \\
 \bottomrule\bottomrule
    \end{tabular}%
    \begin{tablenotes}
    \scriptsize \item \textit{Notes}: In the table, Panel A presents results for money-saving behaviors, with dependent variables being the proportion of transactions with discounts (Columns 1–3), aggregate discount rate (Columns 4–6), and average transaction-level discount rate (Columns 7–9). Panel B examines consumption regularity, with dependent variables measuring month-to-month volatility in shopping patterns across hours of day (Columns 1–3), days of week (Columns 4–6), and ten-day periods of the month (Columns 7–9). Higher (lower) volatility indicates lower (higher) consumption regularity. Controls includes IQ, Big Five personality traits, female, family size, dummies for medium and high income, dummies for medium and high education, and dummies for medium and elder age. Tables~\ref{tab:money-saving} and \ref{tab:volatility_amt} report the results that include the coefficients of control variables. Robust standard errors are in parentheses. {*}~\(p<0.10\), {**}~\(p<0.05\), {***}~\(p<0.01\).
      \end{tablenotes}
    \end{threeparttable}}
\end{table}

Conversely, consistency measure from the supermarket predicts consumption regularity (Table~\ref{tab:predictive_validity}, Panel~B). Consumers with higher consistency scores in the scanner data exhibit shopping routines with significantly less fluctuations, in both 2019–2020 and 2021. Their transaction amount in the supermarket is more stable in terms of hours and days across months, showing significantly less month-to-month volatility. Specifically, a one-standard-deviation increase in CCEI scores is associated with an 8.3–12.7\% standard-deviation reduction in the month-to-month volatility measured using hours of day, a 7.7–9.5\% standard-deviation reduction using days of week, and a 6.0–10.6\% standard-deviation reduction using ten-day periods of the month. The results remain robust when using transaction counts instead of transaction amount to measure volatility (Table~\ref{tab:volatility_num}). Notably, none of the consistency measures from the experiment show significant correlations with any of these indicators of consumption regularity.

Overall, consistency in both the experimental and scanner data demonstrates predictive validity and correlates with distinct, economically meaningful consumer behaviors. This finding alleviates concerns that these measures may simply capture noise. Instead, it provides evidence in support of the multidimensional nature of choice consistency.

\section{Concluding Remarks}
\label{sec:conclusion}
Economic analysis posits that decision makers aim to maximize their utility within budget constraints, and revealed preference analysis provides a framework to evaluate the consistency in choice data. Moreover, consistency score is proposed as a measure of decision quality and is linked to wealth disparities, development gaps, and policy effectiveness. Building on this research, we conduct a comprehensive study to explore the generalizability of consistency scores by integrating both experimental data and scanner data. Our primary findings reveal a lack of correlation between individuals' consistency scores for risky decisions made in the experiment and food consumption choices observed in supermarkets. Within the experimental setting, we observe on average moderate correlations among different tasks using budgetary interface; similarly, in the supermarket context, food consumption decisions across categories and years also show significantly positive correlations. We further show that these correlational patterns remain robust in both settings when using a wide range of consistency indices proposed in the literature. Our study adds to the understanding that preferences and consistency derived from observable choices do not necessarily generalize across different environments. 

We explore several unobservable factors that may underpin consistency scores in different environments, including randomness in choice behaviors, budget constraints, formation of preferences, and heuristic rules. In line with this, \citet{choi2014more} highlight the complexities of measuring decision quality and argue that assessing decision quality in real world contexts is challenging because decision makers ``might have different preferences over the same outcomes, or face different but unobserved incentives and constraints, or have different information or hold different beliefs''. To address these issues, they propose the use of choice consistency as a measure of decision-making quality, particularly in relation to departures from GARP using experimental data. Their study, along with several subsequent ones, further demonstrates that the consistency measure holds real-world significance even after accounting for unobserved constraints, preferences, and beliefs. Here, we present additional results to pin down some underlying factors of consistency measures. In experimental data, we find that consistency is shaped by learning effects, heuristic rules of choosing the middle option and responding to price changes, and the cognitive ability of individuals. In scanner data, consistency is influenced by contextual factors such as seasonality, time of day, promotional discounts, as well as individual shopping experiences. 

In general, these findings suggest that the observed lack of correlation in consistency scores between the two settings may stem in part from measurement errors, particularly in how budget constraints are constructed, as well as the influence of contextual factors. This disconnect may also be due in part to the multidimensional nature of the quality of decision-making. Specifically, participants in the experiment are often faced with novel and abstract problems, while consumers in supermarkets typically make repeated daily decisions in familiar environments. To support this interpretation, we present additional evidence that consistency in both the experimental and scanner data effectively predicts distinct consumer behaviors.
This distinction closely resembles the difference between the decision-making capacity of a routine task and a problem-solving task \citep{deming2021growing, cerigioni2021dual,ilut2023economic}. It is also reflected in the contrast between fluid intelligence, which is the ability to reason and solve problems in unfamiliar situations, and crystallized intelligence, which involves accumulated knowledge, skills, and experiences \citep{cattell1943measurement}.

In summary, our study contributes to a fundamental aspect of economics: choice consistency within the framework of utility maximization under budget constraints \citep{samuelson1938note, afriat1967construction,varian1982nonparametric}. It highlights a disconnect between experimental settings and supermarket environments, moderate correlations within experimental tasks, and weak to moderate correlations in supermarket settings. Our results suggest that the revealed preference measures have yet to capture the intricate and multidimensional nature of decision-making. Looking ahead, future research may explore alternative measures and frameworks of choice consistency that can more accurately capture the complexities of decision-making in controlled laboratory experiments, field experiments, and everyday life.

\clearpage
\singlespacing
\bibliographystyle{aea}
\bibliography{main.bib}

\onehalfspacing
\newpage
\appendix
\renewcommand\theHtable{Appendix.\thetable}

\begin{center}
    \LARGE{\textbf{Online Appendices}}
\end{center}

\setlength\parindent{12pt}

\section{Additional Figures and Tables}
\label{app:results}

\setcounter{table}{0}
\renewcommand{\thetable}{A\arabic{table}}
\setcounter{figure}{0}
\renewcommand{\thefigure}{A\arabic{figure}}

\begin{figure}[H]
\centering

\includegraphics[width=\textwidth]{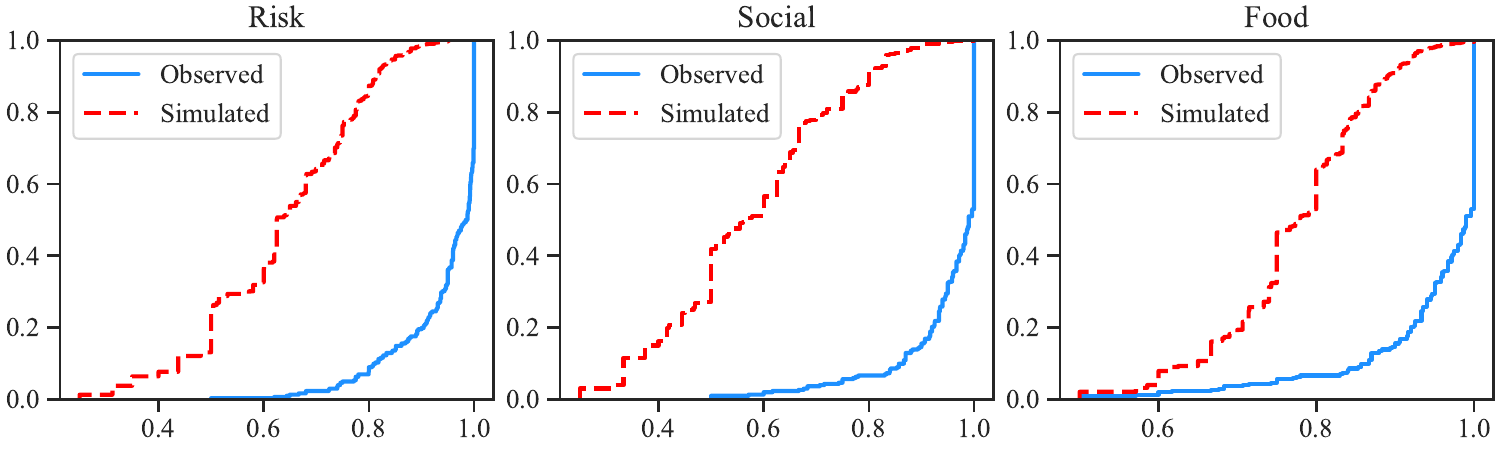}

\caption{Distribution of CCEI Scores in Experiment 2}\label{Rationality Index between Domains}

\end{figure}

\begin{figure}[H]

    \centering
    \includegraphics[width=0.9\textwidth]{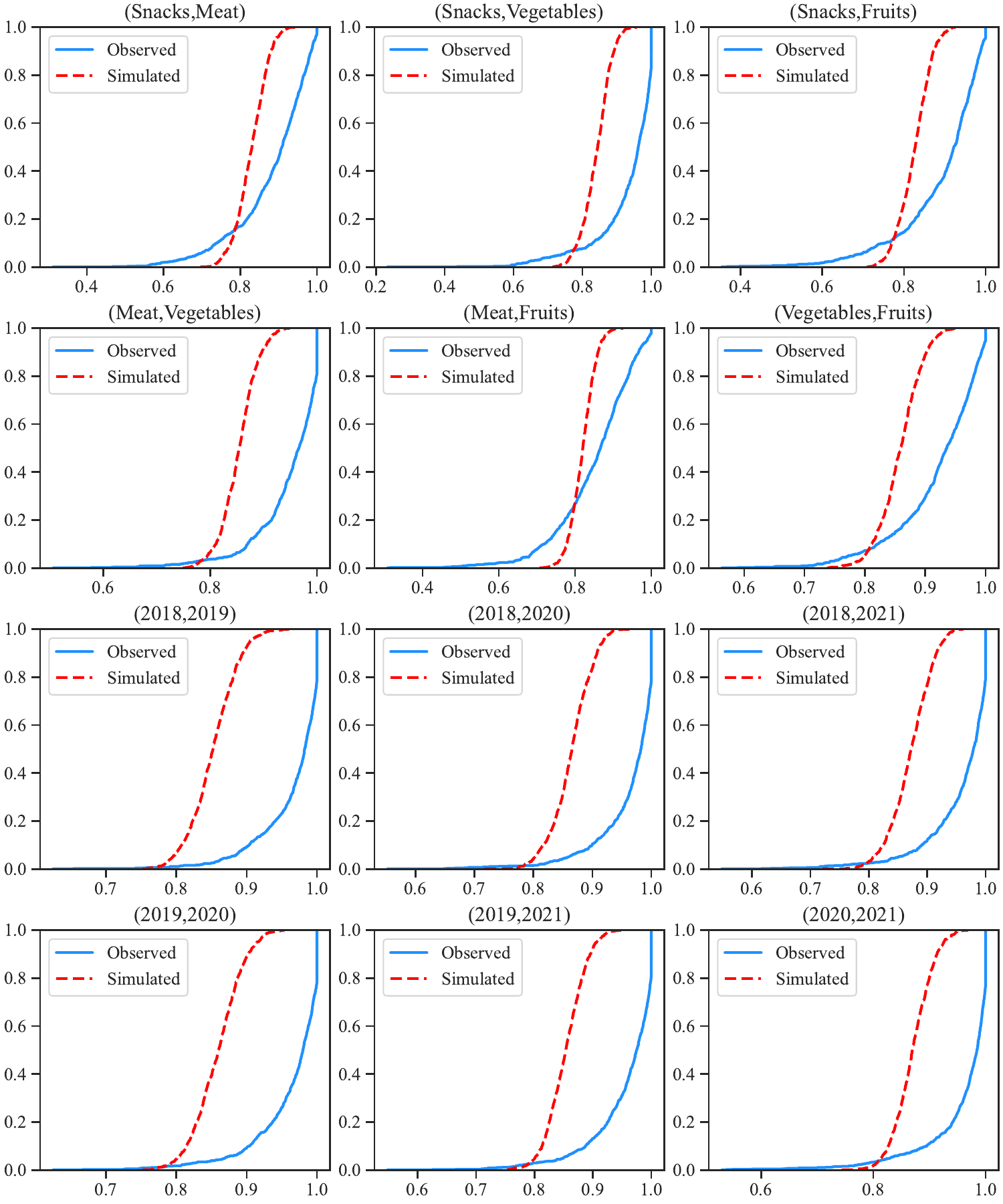}
    
    \caption{Distribution of CCEI Scores in Scanner Dataset 2}\label{fig:study3_des}
    
\end{figure}

\newpage

\begin{table}[H]
\centering
\caption {\centering Summary Statistics of Participants in Two Experiments}\label{tab:demo}
\begin{threeparttable}
\begin{tabular}{llcc}
\toprule \toprule
\textit{Panel A: Demographics} & & & \\
\midrule
\multirow{2}{*}{Variable} & \multirow{2}{*}{Category} & \multicolumn{2}{c}{Percent} \\
\cmidrule{3-4} 
& & Experiment 1 & Experiment 2 \\
\midrule
\multirow{2}{*}{Gender} & Female & 79.9\% & 64.6\% \\
 & Male & 20.1\% & 35.4\% \\
 \midrule
\multirow{3}{*}{Age} & Before 1969 & 19.5\% & 18.5\% \\
 & in 1970-1989 & 75.0\% & 50.7\% \\
 &  After 1990 & 5.5\% & 30.8\% \\
 \midrule
\multirow{3}{*}{Family Monthly Income} & $<$ RMB 5,000 & 24.4\% & 19.2\% \\
 & RMB 5,000-10,000 & 40.0\% & 29.8\% \\
 & $>$ RMB 10,000 & 35.6\% & 51.0\% \\
 \midrule
\multirow{3}{*}{Education} &  $<$ High school & 44.9\% & 36.8\% \\
 & High school & 29.8\% & 19.5\% \\
 & $>$ High school & 25.3\% & 43.7\% \\
 \midrule
\multirow{5}{*}{Family Size} & One & 1.7\% & 2.3\% \\
 & Two & 7.1\% & 10.6\% \\
 & Three & 36.9\% & 35.4\% \\
 & Four & 31.8\% & 32.5\% \\
 & $>$Four & 22.5\% & 19.2\% \\
 \midrule
 N & & 1,055 & 302 \\
 \midrule
\multicolumn{4}{l}{ \textit{Panel B: Consumption behavior (Meat,Veg.) in 2019 and 2020}}\\
 \midrule
\multicolumn{2}{l}{Variable} & Experiment 1 & Experiment 2 \\
\midrule
\multicolumn{2}{l}{\# Months with purchase records} &  24.0 & 8.9 \\
\multicolumn{2}{l}{Monthly expenditure (RMB)} &  482.2 & 245.3 \\
\multicolumn{2}{l}{Monthly shopping days}   &  11.8 & 5.1 \\
\midrule
N & & 1,055 & 117 \\
\bottomrule\bottomrule
\end{tabular}%
		
		\begin{tablenotes}
			\scriptsize \item \textit{Notes}: For participants in Experiment 2, we can match 117 of them who have consumption records of both meat and vegetables in the supermarket in at least one month.
		\end{tablenotes}
	\end{threeparttable}
\end{table}

\newpage 

\begin{table}[H]
\footnotesize
    \centering
\caption{Consistency Indices in Experiment 1 and Scanner Dataset 1}\label{tab:rationality desc}
\begin{tabular}{lcccccccccccc}
\toprule \toprule
\multirow{2}{*}{} & \multicolumn{6}{c}{Observed} & \multicolumn{6}{c}{Simulated} \\
\cmidrule(r){2-7} \cmidrule(l){8-13}
 & N & Mean & SD & Min & Median & Max & N & Mean & SD & Min & Median & Max \\
 \midrule
\multicolumn{13}{l}{\textit{Panel A: Risk task in Experiment 1}} \\
 \midrule
 CCEI & 1,055 & 0.941 & 0.094 & 0.350 & 0.988 & 1 & 1,000 & 0.635 & 0.148 & 0.250 & 0.625 & 0.952 \\
HMI & 1,055 & 0.883 & 0.094 & 0.591 & 0.909 & 1 & 1,000 & 0.697 & 0.074 & 0.455 & 0.682 & 0.909 \\
MPI & 1,055 & 0.045 & 0.060 & 0 & 0.020 & 0.415 & 1,000 & 0.191 & 0.051 & 0.048 & 0.192 & 0.355 \\
MCI & 1,055 & 0.010 & 0.028 & 0 & 0.001 & 0.435 & 1,000 & 0.125 & 0.077 & 0.005 & 0.112 & 0.623 \\
 \midrule
\multicolumn{13}{l}{\textit{Panel B: Consumption (Meat,Veg.) in Scanner Dataset 1}} \\
 \midrule
CCEI & 1,055 & 0.946 & 0.063 & 0.597 & 0.965 & 1 & 1,055 & 0.847 & 0.034 & 0.750 & 0.847 & 0.953 \\
HMI & 1,055 & 0.925 & 0.053 & 0.708 & 0.917 & 1 & 1,055 & 0.860 & 0.026 & 0.771 & 0.860 & 0.932 \\
MPI & 1,055 & 0.056 & 0.050 & 0 & 0.045 & 0.294 & 1,055 & 0.125 & 0.022 & 0.050 & 0.125 & 0.203 \\
MCI & 1,055 & 0.004 & 0.006 & 0 & 0.002 & 0.049 & 1,055 & 0.018 & 0.008 & 0.002 & 0.017 & 0.065   \\
    \bottomrule    \bottomrule
\end{tabular}
\end{table}

\begin{table}[H]
\footnotesize
    \centering
\caption{CCEI Scores in Experiment 2}\label{tab:rationality desc2}
\begin{tabular}{lcccccccccccc}
\toprule \toprule
\multirow{2}{*}{} & \multicolumn{6}{c}{Observed} & \multicolumn{6}{c}{Simulated} \\
\cmidrule(r){2-7} \cmidrule(l){8-13}
 & N & Mean & SD & Min & Median & Max & N & Mean & SD & Min & Median & Max \\
 \midrule
Risk & 302 & 0.943 & 0.084 & 0.500 & 0.986 & 1 & 1,000 & 0.635 & 0.148 & 0.250 & 0.625 & 0.952 \\
Social & 302 & 0.927 & 0.138 & 0.250 & 1 & 1 & 1,000 & 0.572 & 0.162 & 0.250 & 0.562 & 1 \\
Food & 302 & 0.950 & 0.089 & 0.500 & 0.990 & 1 & 1,000 & 0.773 & 0.096 & 0.500 & 0.775 & 1 \\
    \bottomrule    \bottomrule
\end{tabular}
\end{table}

\newpage

\begin{table}[H]
\footnotesize
    \centering
\caption{CCEI Scores in Scanner Dataset 2}\label{tab:rationality desc3}
\hspace*{-0.5cm}
\begin{tabular}{lcccccccccccc}
\toprule \toprule
 & \multicolumn{6}{c}{Observed} & \multicolumn{6}{c}{Simulated} \\
\cmidrule(r){2-7} \cmidrule(l){8-13}
\multicolumn{1}{c}{} & N & Mean & SD & Min & Median & Max & N & Mean & SD & Min & Median & Max \\
\midrule
\multicolumn{13}{l}{\textit{Panel A: CCEI across categories}} \\
\midrule(Snacks,Meat) & 822 & 0.883 & 0.099 & 0.312 & 0.909 & 1 & 822 & 0.829 & 0.041 & 0.698 & 0.831 & 0.945 \\
(Snacks,Veg.) & 822 & 0.933 & 0.090 & 0.234 & 0.962 & 1 & 822 & 0.841 & 0.041 & 0.715 & 0.844 & 0.956 \\
(Snacks,Fruits) & 822 & 0.892 & 0.102 & 0.354 & 0.922 & 1 & 822 & 0.825 & 0.039 & 0.711 & 0.827 & 0.924 \\
(Meat,Veg.) & 822 & 0.947 & 0.063 & 0.506 & 0.966 & 1 & 822 & 0.853 & 0.034 & 0.750 & 0.854 & 0.948 \\
(Meat,Fruits) & 822 & 0.850 & 0.101 & 0.308 & 0.867 & 1 & 822 & 0.820 & 0.033 & 0.709 & 0.820 & 0.925 \\
(Veg.,Fruits) & 822 & 0.920 & 0.071 & 0.563 & 0.936 & 1 & 822 & 0.858 & 0.035 & 0.739 & 0.860 & 0.953 \\
\midrule
\multicolumn{13}{l}{\textit{Panel B: CCEI across time periods}} \\
\midrule
(2018,2019) & 938 & 0.966 & 0.046 & 0.625 & 0.982 & 1 & 938 & 0.852 & 0.034 & 0.753 & 0.852 & 0.961 \\
(2018,2020) & 938 & 0.963 & 0.053 & 0.553 & 0.982 & 1 & 938 & 0.864 & 0.035 & 0.718 & 0.865 & 0.970 \\
(2018,2021) & 938 & 0.959 & 0.057 & 0.549 & 0.979 & 1 & 938 & 0.871 & 0.036 & 0.721 & 0.872 & 0.961 \\
(2019,2020) & 938 & 0.963 & 0.049 & 0.625 & 0.980 & 1 & 938 & 0.859 & 0.034 & 0.753 & 0.860 & 0.953 \\
(2019,2021) & 938 & 0.956 & 0.056 & 0.549 & 0.975 & 1 & 938 & 0.853 & 0.033 & 0.754 & 0.854 & 0.950 \\
(2020,2021) & 938 & 0.961 & 0.062 & 0.530 & 0.985 & 1 & 938 & 0.870 & 0.034 & 0.745 & 0.869 & 0.968 \\
\bottomrule \bottomrule
\end{tabular}
\end{table}

\newpage

\begin{table}[H]
\caption{Correlations of Consistency Indices Between Experiment 1 and Scanner Dataset~1} 
\label{tab:mechanism_s1} 
\centering 
\begin{threeparttable} 
\begin{tabular}{p{10em} p{10em} >{\centering\arraybackslash}p{10em}}
\toprule \toprule 
Exp: Risk & Scanner: Consumption & Correlation \\
\midrule
\multicolumn{3}{l}{\textit{Panel A: Consistency indices}} \\
\midrule
CCEI & CCEI & -0.011 \\
HMI & HMI & 0.050 \\
MPI & MPI & -0.006 \\
MCI & MCI & -0.007 \\
\midrule
\multicolumn{3}{l}{\textit{Panel B: Power of the test}} \\
\midrule
CCEI & Selten score & -0.003 \\
CCEI & Power-adjusted CCEI & -0.009 \\
\midrule
\multicolumn{3}{l}{\textit{Panel C: More (realistic) assumptions on preferences}} \\
\midrule
Normal & Normal & -0.016 \\
Homothetic & Homothetic & -0.005 \\
CCEI & Quasilinear & -0.060* \\
FOSD & CCEI & -0.042 \\
EU & CCEI & -0.045 \\
\midrule
\multicolumn{3}{l}{\textit{Panel D: Preference estimation}} \\
\midrule
LL & LL & 0.042 \\
			\bottomrule \bottomrule
		\end{tabular}
		
		\begin{tablenotes}
			\scriptsize \item \textit{Notes}: \( N = 1,055 \). 
   Spearman’s rank correlations are reported.  {*} \(p<0.10\), {**} \(p<0.05\), {***} \(p<0.01\).
		\end{tablenotes}
	\end{threeparttable}
\end{table}

\newpage

\begin{table}[H]
\caption{Correlations of Consistency Indices in Experiment 2} 
\label{tab:mechanism_s2} 
\centering 
\begin{threeparttable} 
\begin{tabular}{p{10em} >{\centering\arraybackslash}p{7em} >{\centering\arraybackslash}p{7em}  >{\centering\arraybackslash}p{7em}}
\toprule \toprule 
 & Risk vs. Social & Risk vs. Food & Social vs. Food \\
\midrule
\multicolumn{4}{l}{\textit{Panel A: Consistency indices}} \\
\midrule
CCEI & 0.412*** & 0.305*** & 0.254*** \\
HMI & 0.449*** & 0.325*** & 0.306*** \\
MPI & 0.408*** & 0.309*** & 0.241*** \\
MCI & 0.408*** & 0.339*** & 0.249*** \\
\midrule
\multicolumn{4}{l}{\textit{Panel B: More (realistic) assumptions on preferences}} \\
\midrule
Normal & 0.426*** & 0.349*** & 0.324*** \\
Homothetic & 0.495*** & 0.283*** & 0.204*** \\
FOSD & 0.260*** & 0.187*** & $-$ \\
EU & 0.263*** & 0.191*** & $-$ \\
\midrule
\multicolumn{4}{l}{\textit{Panel C: Preference estimation}} \\
\midrule
LL & 0.522*** & 0.379*** & 0.383***\\
			\bottomrule \bottomrule
		\end{tabular}
		
		\begin{tablenotes}
			\scriptsize \item \textit{Notes}: $N=302$. Spearman’s rank correlations are reported. Each coefficient is calculated using the index as a measure for both consistency scores. Two exceptions are FOSD and EU, which are applied exclusively to the risk task; for these, we calculate the correlations between them and CCEI scores for other tasks. $-$ denotes that the measure is not applicable. {*}~\(p<0.10\), {**}~\(p<0.05\), {***}~\(p<0.01\).
		\end{tablenotes}
	\end{threeparttable}
\end{table}

\newpage

\begin{table}[H]
\caption{Correlations of Consistency Indices in Scanner Dataset 2} 
\label{tab:mechanism_s3} 
\centering 
\footnotesize 
\hspace*{-1.4cm}
\begin{threeparttable} 
\begin{tabular}{p{9.5em}cccccc}
\toprule \toprule
 & \multicolumn{3}{c}{Across categories ($N=822$)} & \multicolumn{3}{c}{Across time ($N=938$)} \\
\cmidrule(r){2-4}\cmidrule(l){5-7}
 & \scriptsize \makecell[c]{(Snacks,Meat) vs.\\ (Veg.,Fruits) }&\scriptsize  \makecell[c]{(Snacks,Veg.) vs. \\ (Meat,Fruits) }&\scriptsize  \makecell[c]{(Snacks,Fruits) vs. \\ (Meat,Veg.)} & \scriptsize \makecell[c]{(2018,2019) vs.\\ (2020,2021)} &\scriptsize  \makecell[c]{(2018,2020) vs. \\ (2019,2021)} & \scriptsize \makecell[c]{(2018,2021) vs.\\ (2019,2020)} \\
\midrule
\multicolumn{7}{l}{\textit{Panel A: Consistency indices}} \\
\midrule
CCEI & 0.173*** & -0.046 & 0.096*** & 0.112*** & 0.186*** & 0.143*** \\
HMI & 0.061* & -0.016 & 0.100*** & 0.096*** & 0.114*** & 0.132*** \\
MPI & 0.194*** & -0.035 & 0.099*** & 0.142*** & 0.220*** & 0.160*** \\
MCI & 0.154*** & -0.048 & 0.102*** & 0.131*** & 0.192*** & 0.156*** \\
\midrule
\multicolumn{7}{l}{\textit{Panel B: Power of the test}} \\
\midrule
Selten score & 0.098*** & -0.069** & 0.041 & 0.045 & 0.127*** & 0.134*** \\
Power-adjusted CCEI & 0.121*** & -0.070** & 0.016 & 0.047 & 0.127*** & 0.117*** \\
\midrule
\multicolumn{7}{l}{\textit{Panel C: More (realistic) assumptions on preferences}} \\
\midrule
Normal & 0.191*** & 0.214*** & 0.114*** & 0.449*** & 0.502*** & 0.482*** \\
Homothetic & 0.226*** & 0.033 & 0.137*** & 0.218*** & 0.341*** & 0.342*** \\
Quasilinear & 0.265*** & 0.329*** & 0.165*** & 0.260*** & 0.379*** & 0.430*** \\
\midrule
\multicolumn{7}{l}{\textit{Panel D: Preference estimation}} \\
\midrule
LL & 0.290*** & -0.003 & 0.172*** & 0.421*** & 0.529*** & 0.468*** \\
			\bottomrule \bottomrule
		\end{tabular}
		
		\begin{tablenotes}
			\scriptsize \item \textit{Notes}: Spearman’s rank correlations are reported. Each coefficient is calculated using the index as a measure for both consistency scores. {*}~\(p<0.10\), {**}~\(p<0.05\), {***}~\(p<0.01\).
		\end{tablenotes}
	\end{threeparttable}
\end{table}

\newpage

\begin{table}[H]
\centering
\caption{Correlations of Alternative Indices for Budget Constraints and Heuristic Rules}
\label{app:corr_budget_heuristics}
 \begin{threeparttable}
\begin{tabular}{p{14em} >{\centering\arraybackslash}p{6em} >{\centering\arraybackslash}p{6em}  >{\centering\arraybackslash}p{6em}}
\toprule \toprule
 & PMI & GAPP & Downward \\
\midrule
\multicolumn{4}{l}{\textit{Panel A: Experiment   1 vs. Scanner Dataset 1}} \\
\midrule
Risk vs (Veg.,Meat) & 0.023 & -0.025 & -0.029 \\
\midrule
\multicolumn{4}{l}{\textit{Panel B: Experiment   2}} \\
\midrule
Risk vs Social & 0.378*** & -- & 0.539*** \\
Risk vs  Food & 0.220*** & -- & 0.324*** \\
Social vs  Food & 0.208*** & -- & 0.345*** \\
\midrule
\multicolumn{4}{l}{\textit{Panel C: Scanner   Dataset 2}} \\
\midrule
(Snacks,Meat) vs.   (Veg.,Fruits) & 0.151*** & 0.092*** & 0.084** \\
(Snacks,Veg.) vs.   (Meat,Fruits) & 0.014 & -0.009 & 0.042 \\
(Snacks,Fruits) vs.   (Meat,Veg.) & 0.081** & 0.083** & 0.098*** \\
(2018,2019) vs. (2020,2021) & 0.068** & 0.198*** & 0.091*** \\
(2018,2020) vs. (2019,2021) & 0.122*** & 0.235*** & 0.224*** \\
(2018,2021) vs. (2019,2020) & 0.097*** & 0.279*** & 0.199*** \\
\bottomrule \bottomrule
\end{tabular}
		\begin{tablenotes}
			\scriptsize \item \textit{Notes}: Spearman’s rank correlations are reported. $-$ denotes that the measure is not applicable. {*} \(p<0.10\), {**} \(p<0.05\), {***} \(p<0.01\). 
		\end{tablenotes}
\end{threeparttable}
\end{table}

\newpage

\begin{table}[H]
  \centering
\captionsetup{justification=centering}
  \caption{Average Proportions of GARP Violation Across Combinations of Seasons in Scanner Dataset~1}
  \label{tab:season}%
  \begin{threeparttable}
  \begin{tabular}{p{6em} >{\centering\arraybackslash}p{6em} >{\centering\arraybackslash}p{6em}  >{\centering\arraybackslash}p{6em} >{\centering\arraybackslash}p{6em}}
    \toprule   \toprule
 & Spring & Summer & Fall & Winter \\
 \midrule
Spring & 1.04\% & 1.02\% & 1.08\% & 1.03\% \\
Summer &  & 0.87\% & 0.93\% & 0.91\% \\
Fall &  &  & 0.88\% & 0.84\% \\
Winter &  &  &  & 0.85\% \\
    \bottomrule   \bottomrule
    \end{tabular}%
    \begin{tablenotes}
\scriptsize
\item \textit{Notes}: For each combination of seasons, the number in each cell represents the proportion of month pairs that exhibit violations of GARP within all month pairs, averaged across all consumers.
      \end{tablenotes}
    \end{threeparttable}
\end{table}%

\begin{table}[H]
\footnotesize
    \centering
  \caption{Pairwise Comparisons Across Different Combinations of Seasons in Scanner Dataset~1}
  \label{tab:season_compare}%
  \begin{threeparttable}
\begin{tabular}{lccccc}
\toprule \toprule
 & (Spring,Fall) & (Spring,Winter) & (Summer,Fall) & (Summer,Winter) & (Fall,Winter) \\
 \midrule
 (Spring,Summer) & 0.200 & 0.803 & 0.119 & 0.043 &  \\
(Spring,Fall) &  & 0.269 & 0.006 &  & 0.000 \\
(Spring,Winter) &  &  &  & 0.017 & 0.000 \\
(Summer,Fall) &  &  &  & 0.610 & 0.057 \\
(Summer,Winter) &  &  &  &  & 0.140 \\
    \bottomrule   \bottomrule
    \end{tabular}%
    \begin{tablenotes}
\scriptsize
\item \textit{Notes}: The table reports the pairwise comparisons of results in Table \ref{tab:season} between different season combinations. \textit{p}-values of paired $t$-tests are reported.
      \end{tablenotes}
    \end{threeparttable}
\end{table}%

\begin{table}[H]
\captionsetup{justification=centering}
  \caption{Comparisons of Estimated Parameters for Food Preference Across Seasons in Scanner Dataset~1} \label{tab:est_pref season}
    \centering
\begin{threeparttable}
\begin{tabular}{cccccccc}
    \toprule\toprule
 &  & \multicolumn{3}{c}{$\hat{\alpha}$} & \multicolumn{3}{c}{$\hat{\rho}$} \\
 \cmidrule(r){3-5} \cmidrule(l){6-8}
Season: $s_{1}$ & Season: $s_{2}$ & Mean: $s_{1}$ & Mean: $s_{2}$ & $p$-value & Mean: $s_{1}$ & Mean: $s_{2}$ & $p$-value \\
\midrule
Spring & Summer & 0.371 & 0.422 & 0.000 & -6.356 & -5.816 & 0.108 \\
Spring & Fall & 0.371 & 0.445 & 0.000 & -6.356 & -6.303 & 0.886 \\
Spring & Winter & 0.371 & 0.425 & 0.000 & -6.356 & -6.786 & 0.224 \\
Summer & Fall & 0.422 & 0.445 & 0.124 & -5.816 & -6.303 & 0.186 \\
Summer & Winter & 0.422 & 0.425 & 0.861 & -5.816 & -6.786 & 0.009 \\
Fall & Winter & 0.445 & 0.425 & 0.174 & -6.303 & -6.786 & 0.197 \\
 \bottomrule\bottomrule
    \end{tabular}%
    \begin{tablenotes}
    \scriptsize
    \item \textit{Notes}: \textit{p}-values of paired $t$-tests are reported.
      \end{tablenotes}
    \end{threeparttable}
 \end{table}

\begin{table}[H]
\captionsetup{justification=centering}
  \caption{Tobit Regressions for Consistency Indices in Experiment~1 and Scanner Dataset~1} \label{tab:tobit_behavioral}
    \centering
    \footnotesize
    \begin{threeparttable}
\begin{tabular}{p{9em}>{\centering\arraybackslash}p{9em}>{\centering\arraybackslash}p{9em}>{\centering\arraybackslash}p{9em}}
    \toprule\toprule
                           & Exp: CCEI & Exp: FOSD & Scanner: CCEI \\
                           & (1)       & (2)       & (3)           \\
\midrule
\textit{IQ}                & 0.003     & 0.014***  & 0.000         \\
\textit{}                  & (0.002)   & (0.005)   & (0.001)       \\
\textit{Frequency}         & 0.065     & 0.036     & 0.272***      \\
\textit{}                  & (0.097)   & (0.168)   & (0.049)       \\
\textit{Conscientiousness} & 0.001     & 0.002     & -0.002*       \\
\textit{}                  & (0.002)   & (0.004)   & (0.001)       \\
\textit{Extraversion}      & 0.001     & -0.001    & 0.001         \\
\textit{}                  & (0.002)   & (0.004)   & (0.001)       \\
\textit{Agreeableness}     & -0.004*   & -0.006    & 0.001         \\
\textit{}                  & (0.002)   & (0.004)   & (0.001)       \\
\textit{Openness}          & -0.004*   & -0.006    & 0.002         \\
\textit{}                  & (0.002)   & (0.004)   & (0.001)       \\
\textit{Neuroticism}       & -0.007*** & -0.009**  & -0.000        \\
\textit{}                  & (0.002)   & (0.004)   & (0.001)       \\
\textit{Female}            & -0.003    & 0.027     & -0.011*       \\
\textit{}                  & (0.010)   & (0.019)   & (0.006)       \\
\textit{Medium Income}     & -0.006    & 0.011     & 0.001         \\
\textit{}                  & (0.010)   & (0.019)   & (0.005)       \\
\textit{High income}       & 0.017     & 0.046**   & -0.001        \\
\textit{}                  & (0.012)   & (0.020)   & (0.006)       \\
\textit{Medium education}  & -0.003    & -0.005    & -0.003        \\
\textit{}                  & (0.010)   & (0.017)   & (0.005)       \\
\textit{High education}    & 0.011     & 0.030     & -0.009        \\
\textit{}                  & (0.011)   & (0.019)   & (0.006)       \\
\textit{Medium age}        & 0.022     & 0.063*    & -0.005        \\
\textit{}                  & (0.020)   & (0.036)   & (0.010)       \\
\textit{Elder age}         & 0.018     & 0.068*    & -0.006        \\
\textit{}                  & (0.022)   & (0.038)   & (0.010)       \\
\textit{Family size}       & -0.001    & -0.004    & 0.001         \\
\textit{}                  & (0.004)   & (0.007)   & (0.002)       \\
\textit{Simulated CCEI}    &           &           & 0.503***      \\
                           &           &           & (0.062)       \\
\textit{Constant}          & 0.924***  & 0.680***  & 0.507***      \\
                           & (0.032)   & (0.058)   & (0.055)       \\
 \midrule
N                          & 1,055     & 1,055     & 1,055         \\
 \bottomrule\bottomrule
    \end{tabular}%
\begin{tablenotes}
\scriptsize
\item \textit{Notes}: The upper limit for Tobit regressions is 1. \textit{IQ} is the number of correct answers in a seven-question version of Raven’s Progressive Matrices. \textit{Frequency} is the average monthly shopping days (divided by 100) for meat and vegetables from 2019 to 2020. Dummy for medium (high) income indicates family income between RMB 5,001 and 10,000 (more than RMB 10,000) per month. Dummy for medium (high) education indicates education degree of high school (above high school). Dummy for medium (elder) age refers to those born between 1970 and 1989 (before 1969). In the scanner dataset, simulated CCEI is added as an additional control to address the heterogeneous power issue. Robust standard errors are in parentheses. {*}~\(p<0.10\), {**}~\(p<0.05\), {***}~\(p<0.01\).
      \end{tablenotes}
    \end{threeparttable}
\end{table}

\newpage

\begin{table}[H]
\captionsetup{justification=centering}
  \caption{OLS Regressions for Relative Ranking of Consistency Indices in Experiment~1 and Scanner Dataset~1} \label{tab:rank_behavioral}
    \centering
    \footnotesize
    \begin{threeparttable}
\begin{tabular}{p{9em}>{\centering\arraybackslash}p{9em}>{\centering\arraybackslash}p{9em}>{\centering\arraybackslash}p{9em}}
    \toprule\toprule
                           & Exp: CCEI & Exp: FOSD & Scanner: CCEI \\
                           & (1)       & (2)       & (3)           \\
\midrule
\textit{IQ}                & 0.005     & 0.015***  & 0.005         \\
\textit{}                  & (0.006)   & (0.006)   & (0.005)       \\
\textit{Frequency}         & 0.121     & 0.040     & 1.158***      \\
\textit{}                  & (0.254)   & (0.221)   & (0.220)       \\
\textit{Conscientiousness} & -0.004    & -0.002    & -0.010*       \\
\textit{}                  & (0.006)   & (0.005)   & (0.005)       \\
\textit{Extraversion}      & 0.003     & 0.001     & 0.000         \\
\textit{}                  & (0.006)   & (0.005)   & (0.005)       \\
\textit{Agreeableness}     & -0.008    & -0.006    & 0.005         \\
\textit{}                  & (0.006)   & (0.005)   & (0.005)       \\
\textit{Openness}          & -0.008    & -0.006    & 0.004         \\
\textit{}                  & (0.006)   & (0.005)   & (0.005)       \\
\textit{Neuroticism}       & -0.017*** & -0.014*** & -0.003        \\
\textit{}                  & (0.006)   & (0.005)   & (0.005)       \\
\textit{Female}            & -0.012    & 0.013     & -0.033        \\
\textit{}                  & (0.028)   & (0.025)   & (0.024)       \\
\textit{Medium Income}     & -0.023    & 0.005     & 0.016         \\
\textit{}                  & (0.027)   & (0.024)   & (0.023)       \\
\textit{High income}       & 0.047     & 0.053**   & 0.004         \\
\textit{}                  & (0.031)   & (0.027)   & (0.026)       \\
\textit{Medium education}  & 0.023     & -0.001    & -0.013        \\
\textit{}                  & (0.025)   & (0.022)   & (0.022)       \\
\textit{High education}    & 0.032     & 0.042     & -0.044*       \\
\textit{}                  & (0.031)   & (0.027)   & (0.026)       \\
\textit{Medium age}        & 0.045     & 0.068     & -0.031        \\
\textit{}                  & (0.048)   & (0.043)   & (0.042)       \\
\textit{Elder age}         & 0.034     & 0.068     & -0.029        \\
\textit{}                  & (0.053)   & (0.047)   & (0.045)       \\
\textit{Family size}       & -0.001    & -0.011    & -0.000        \\
\textit{}                  & (0.011)   & (0.010)   & (0.010)       \\
\textit{Simulated CCEI}    &           &           & 2.407***      \\
                           &           &           & (0.256)       \\
\textit{Constant}          & 0.474***  & 0.386***  & -1.625***     \\
                           & (0.081)   & (0.074)   & (0.225)       \\
 \midrule
N                          & 1,055     & 1,055     & 1,055         \\
\textit{R$^2$}                      & 0.027     & 0.033     & 0.108         \\
 \bottomrule\bottomrule
    \end{tabular}%
\begin{tablenotes}
\scriptsize
\item \textit{Notes}: Dependent variables are $Relative \ ranking_n = (N-rank(consistency_n) +1) / N$, where n denotes subject, $N$ is the number of observations, and $rank(consistency_n)$ is the rank of subject $n$'s consistency index in descending order. 
\textit{IQ} is the number of correct answers in a seven-question version of Raven’s Progressive Matrices. \textit{Frequency} is the average monthly shopping days (divided by 100) for meat and vegetables from 2019 to 2020. Dummy for medium (high) income indicates family income between RMB 5,001 and 10,000 (more than RMB 10,000) per month. Dummy for medium (high) education indicates education degree of high school (above high school). Dummy for medium (elder) age refers to those born between 1970 and 1989 (before 1969). 
In the scanner dataset, simulated CCEI is added as an additional control to address the heterogeneous power issue. Robust standard errors are in parentheses. {*}~\(p<0.10\), {**}~\(p<0.05\), {***}~\(p<0.01\).
      \end{tablenotes}
    \end{threeparttable}
\end{table}

\newpage

\begin{table}[H]
  \caption{Correlations of Estimated Parameters in Experiment~2} \label{tab: corr_para}
    \centering
    \begin{threeparttable}
\begin{tabular}{p{2em} >{\centering\arraybackslash}p{9em} >{\centering\arraybackslash}p{9em} >{\centering\arraybackslash}p{9em}}
\toprule \toprule
 & Risk vs Social & Risk vs  Food & Social vs  Food \\
 \midrule
$\hat{\alpha}$ & 0.150*** & 0.109* & 0.067  \\
$\hat{\rho}$ & 0.366*** & 0.081    & 0.160*** \\
\bottomrule \bottomrule 
\end{tabular}
\begin{tablenotes}
\scriptsize
\item \textit{Notes}: Spearman’s rank correlations are reported. {*} \(p<0.10\), {**} \(p<0.05\), {***} \(p<0.01\).
      \end{tablenotes}
    \end{threeparttable}
\end{table}

\newpage

\begin{table}[H]
\begin{adjustwidth}{-1.5cm}{-1.5cm} 
\captionsetup{justification=centering}
  \caption{OLS Regressions for Money-saving Behavior}
  \label{tab:money-saving}
    \centering
    \footnotesize
\begin{threeparttable}
\begin{tabular}{lccccccccc}
\toprule\toprule
 & \multicolumn{3}{c}{\makecell[c]{Proportion of transactions \\ with discounts}} & \multicolumn{3}{c}{Aggregate discount rate} & \multicolumn{3}{c}{\makecell[c]{Average transaction-level \\ discount rate}} \\
\cmidrule(lr){2-4}\cmidrule(lr){5-7}\cmidrule(lr){8-10}
 & 2019 & 2020 & 2021 & 2019 & 2020 & 2021 & 2019 & 2020 & 2021 \\
 & (1) & (2) & (3) & (4) & (5) & (6) & (7) & (8) & (9) \\
\midrule
\textit{Exp: FOSD} & 0.045** & 0.043*** & 0.053*** & 0.015 & 0.027** & 0.020* & 0.019* & 0.021** & 0.029*** \\
\textit{} & (0.019) & (0.017) & (0.020) & (0.012) & (0.011) & (0.011) & (0.011) & (0.011) & (0.011) \\
\textit{Scanner: CCEI} & -0.062 & -0.045 & -0.046 & -0.072* & -0.049 & 0.020 & -0.048 & -0.026 & -0.040 \\
\textit{} & (0.066) & (0.059) & (0.069) & (0.039) & (0.035) & (0.036) & (0.037) & (0.035) & (0.040) \\
\textit{IQ} & 0.006*** & 0.007*** & 0.009*** & 0.004** & 0.003** & 0.001 & 0.004*** & 0.004*** & 0.003** \\
 & (0.002) & (0.002) & (0.003) & (0.001) & (0.001) & (0.002) & (0.001) & (0.001) & (0.001) \\
\textit{Conscientiousness} & 0.003* & 0.004* & 0.001 & 0.002 & 0.002* & 0.002 & 0.003** & 0.002* & 0.002 \\
\textit{} & (0.002) & (0.002) & (0.002) & (0.001) & (0.001) & (0.001) & (0.001) & (0.001) & (0.001) \\
\textit{Extraversion} & 0.001 & 0.003* & 0.003 & 0.002 & 0.001 & 0.002 & 0.002 & 0.003** & 0.002 \\
\textit{} & (0.002) & (0.002) & (0.002) & (0.001) & (0.001) & (0.001) & (0.001) & (0.001) & (0.001) \\
\textit{Agreeableness} & 0.004* & 0.002 & 0.006** & 0.003** & 0.002 & 0.003* & 0.003** & 0.001 & 0.003** \\
\textit{} & (0.002) & (0.002) & (0.003) & (0.001) & (0.001) & (0.002) & (0.001) & (0.001) & (0.002) \\
\textit{Openness} & 0.000 & -0.002 & -0.004 & 0.001 & -0.001 & -0.001 & 0.000 & -0.001 & -0.001 \\
\textit{} & (0.002) & (0.002) & (0.003) & (0.001) & (0.001) & (0.001) & (0.001) & (0.001) & (0.002) \\
\textit{Neuroticism} & 0.000 & 0.003 & 0.001 & -0.001 & 0.000 & 0.000 & -0.000 & 0.001 & -0.000 \\
\textit{} & (0.002) & (0.002) & (0.003) & (0.002) & (0.001) & (0.002) & (0.002) & (0.001) & (0.002) \\
\textit{Female} & 0.043*** & 0.018* & 0.014 & 0.029*** & 0.008 & 0.003 & 0.027*** & 0.010* & 0.006 \\
\textit{} & (0.010) & (0.009) & (0.011) & (0.006) & (0.006) & (0.007) & (0.005) & (0.006) & (0.006) \\
\textit{Medium Income} & -0.013 & -0.012 & -0.014 & -0.009 & -0.008 & -0.010 & -0.014** & -0.011* & -0.014** \\
\textit{} & (0.010) & (0.010) & (0.011) & (0.007) & (0.006) & (0.007) & (0.006) & (0.006) & (0.007) \\
\textit{High income} & -0.022** & -0.013 & -0.022* & -0.017** & -0.009 & -0.015** & -0.018*** & -0.011 & -0.017** \\
\textit{} & (0.011) & (0.012) & (0.013) & (0.007) & (0.007) & (0.008) & (0.007) & (0.007) & (0.008) \\
\textit{Medium education} & -0.033*** & -0.040*** & -0.030*** & -0.022*** & -0.021*** & -0.016*** & -0.022*** & -0.023*** & -0.020*** \\
\textit{} & (0.009) & (0.009) & (0.010) & (0.006) & (0.006) & (0.006) & (0.005) & (0.005) & (0.006) \\
\textit{High education} & -0.041*** & -0.042*** & -0.016 & -0.033*** & -0.030*** & -0.016** & -0.031*** & -0.029*** & -0.017** \\
\textit{} & (0.010) & (0.011) & (0.013) & (0.007) & (0.007) & (0.008) & (0.007) & (0.006) & (0.008) \\
\textit{Medium age} & 0.042*** & 0.021 & 0.031* & 0.022*** & 0.009 & 0.012 & 0.025*** & 0.010 & 0.012 \\
\textit{} & (0.013) & (0.015) & (0.019) & (0.009) & (0.010) & (0.011) & (0.007) & (0.009) & (0.010) \\
\textit{Elder age} & 0.051*** & 0.044** & 0.051** & 0.016 & 0.005 & 0.007 & 0.024*** & 0.010 & 0.008 \\
\textit{} & (0.015) & (0.017) & (0.022) & (0.011) & (0.012) & (0.013) & (0.009) & (0.011) & (0.012) \\
\textit{Family size} & -0.000 & -0.000 & -0.003 & -0.001 & -0.001 & 0.002 & -0.002 & -0.001 & -0.002 \\
\textit{} & (0.004) & (0.004) & (0.005) & (0.003) & (0.002) & (0.003) & (0.003) & (0.002) & (0.003) \\
\textit{Constant} & 0.462*** & 0.435*** & 0.439*** & 0.202*** & 0.163*** & 0.115*** & 0.123*** & 0.106*** & 0.127*** \\
 & (0.072) & (0.064) & (0.076) & (0.044) & (0.038) & (0.041) & (0.041) & (0.038) & (0.044) \\
\midrule
N & 1,052 & 1,046 & 1,038 & 1,052 & 1,046 & 1,038 & 1,052 & 1,046 & 1,038 \\
\textit{R$^2$} & 0.101 & 0.069 & 0.051 & 0.112 & 0.064 & 0.041 & 0.121 & 0.068 & 0.054 \\
\bottomrule\bottomrule
    \end{tabular}%
    \begin{tablenotes}
    \scriptsize \item \textit{Notes}: The table presents results for money-saving behaviors, with dependent variables being the proportion of transactions with discounts (Columns 1–3), aggregate discount rate (Columns 4–6), and average transaction-level discount rate (Columns 7–9). \textit{IQ} is the number of correct answers in a seven-question version of Raven’s Progressive Matrices. Dummy for medium (high) education indicates education degree of high school (above high school). Dummy for medium (elder) age refers to those born between 1970 and 1989 (before 1969).  Robust standard errors are in parentheses. {*}~\(p<0.10\), {**}~\(p<0.05\), {***}~\(p<0.01\).
      \end{tablenotes}
    \end{threeparttable}
\end{adjustwidth} 
\end{table}

\newpage

\begin{table}[H]
\begin{adjustwidth}{-1.5cm}{-1.5cm}
\captionsetup{justification=centering}
  \caption{OLS Regressions for Consumption Regularity (Transaction Amount)}
  \label{tab:volatility_amt}
    \centering
    \footnotesize
\begin{threeparttable}
\begin{tabular}{lccccccccc}
\toprule\toprule
 & \multicolumn{3}{c}{Hours of day} & \multicolumn{3}{c}{Days of week} & \multicolumn{3}{c}{Ten-day periods of the month} \\
\cmidrule(lr){2-4}\cmidrule(lr){5-7}\cmidrule(lr){8-10}
 & 2019 & 2020 & 2021 & 2019 & 2020 & 2021 & 2019 & 2020 & 2021 \\
 & (1) & (2) & (3) & (4) & (5) & (6) & (7) & (8) & (9) \\
 \midrule
\textit{Exp: FOSD} & 0.003 & 0.001 & 0.000 & 0.008 & -0.001 & -0.013 & 0.005 & 0.018 & -0.000 \\
\textit{} & (0.003) & (0.004) & (0.004) & (0.007) & (0.009) & (0.009) & (0.010) & (0.015) & (0.017) \\
\textit{Scanner: CCEI} & -0.043*** & -0.039*** & -0.035*** & -0.072*** & -0.075** & -0.077*** & -0.124*** & -0.116** & -0.101** \\
\textit{} & (0.010) & (0.012) & (0.012) & (0.023) & (0.029) & (0.027) & (0.036) & (0.049) & (0.050) \\
\textit{IQ} & -0.000 & -0.001 & -0.001 & -0.000 & -0.001 & -0.001 & 0.001 & -0.000 & 0.001 \\
 & (0.000) & (0.000) & (0.000) & (0.001) & (0.001) & (0.001) & (0.001) & (0.002) & (0.002) \\
\textit{Conscientiousness} & -0.001* & 0.000 & -0.000 & -0.001 & -0.000 & -0.000 & -0.001 & 0.001 & -0.003 \\
\textit{} & (0.000) & (0.000) & (0.000) & (0.001) & (0.001) & (0.001) & (0.001) & (0.002) & (0.002) \\
\textit{Extraversion} & 0.000 & -0.000 & 0.000 & 0.000 & -0.000 & -0.001 & -0.000 & -0.002 & -0.001 \\
\textit{} & (0.000) & (0.000) & (0.000) & (0.001) & (0.001) & (0.001) & (0.001) & (0.002) & (0.002) \\
\textit{Agreeableness} & 0.000 & 0.001 & -0.000 & 0.001 & 0.001 & -0.001 & 0.001 & 0.002 & 0.002 \\
\textit{} & (0.000) & (0.000) & (0.000) & (0.001) & (0.001) & (0.001) & (0.001) & (0.002) & (0.002) \\
\textit{Openness} & 0.001*** & 0.001* & 0.001* & 0.002* & 0.002** & 0.003** & 0.003** & 0.003 & 0.005*** \\
\textit{} & (0.000) & (0.000) & (0.000) & (0.001) & (0.001) & (0.001) & (0.001) & (0.002) & (0.002) \\
\textit{Neuroticism} & 0.000 & 0.000 & -0.000 & 0.000 & 0.000 & -0.000 & -0.000 & 0.000 & -0.001 \\
\textit{} & (0.000) & (0.000) & (0.000) & (0.001) & (0.001) & (0.001) & (0.001) & (0.002) & (0.002) \\
\textit{Female} & 0.005*** & 0.009*** & 0.007*** & 0.008* & 0.016*** & 0.015*** & 0.018*** & 0.027*** & 0.024*** \\
\textit{} & (0.002) & (0.002) & (0.002) & (0.004) & (0.005) & (0.005) & (0.006) & (0.009) & (0.009) \\
\textit{Medium Income} & 0.000 & -0.001 & -0.001 & 0.001 & -0.004 & -0.003 & 0.004 & -0.008 & 0.008 \\
\textit{} & (0.002) & (0.002) & (0.002) & (0.004) & (0.005) & (0.005) & (0.006) & (0.008) & (0.009) \\
\textit{High income} & 0.003* & -0.001 & -0.001 & 0.008* & -0.001 & -0.000 & 0.011* & -0.002 & 0.002 \\
\textit{} & (0.002) & (0.002) & (0.002) & (0.004) & (0.005) & (0.006) & (0.007) & (0.009) & (0.009) \\
\textit{Medium education} & -0.004*** & -0.006*** & -0.006*** & -0.005 & -0.012*** & -0.013*** & -0.011** & -0.019** & -0.020** \\
\textit{} & (0.002) & (0.002) & (0.002) & (0.004) & (0.005) & (0.005) & (0.005) & (0.008) & (0.008) \\
\textit{High education} & -0.002 & -0.004* & -0.003 & -0.001 & -0.006 & -0.003 & -0.010 & -0.009 & -0.006 \\
\textit{} & (0.002) & (0.002) & (0.002) & (0.004) & (0.005) & (0.006) & (0.007) & (0.009) & (0.009) \\
\textit{Medium age} & -0.000 & 0.000 & 0.004 & 0.002 & 0.008 & 0.010 & 0.000 & 0.014 & 0.012 \\
\textit{} & (0.003) & (0.003) & (0.003) & (0.007) & (0.008) & (0.009) & (0.011) & (0.014) & (0.015) \\
\textit{Elder age} & -0.008** & -0.010*** & -0.006 & -0.012 & -0.010 & -0.009 & -0.028** & -0.013 & -0.020 \\
\textit{} & (0.004) & (0.004) & (0.004) & (0.008) & (0.009) & (0.009) & (0.012) & (0.016) & (0.016) \\
\textit{Family size} & -0.002*** & -0.000 & -0.000 & -0.005*** & -0.002 & 0.000 & -0.007*** & -0.002 & -0.006* \\
\textit{} & (0.001) & (0.001) & (0.001) & (0.002) & (0.002) & (0.002) & (0.002) & (0.003) & (0.003) \\
\textit{Constant} & 0.115*** & 0.118*** & 0.120*** & 0.212*** & 0.241*** & 0.265*** & 0.329*** & 0.333*** & 0.385*** \\
 & (0.012) & (0.013) & (0.014) & (0.025) & (0.032) & (0.030) & (0.040) & (0.054) & (0.054) \\
\midrule
N & 1,050 & 1,043 & 1,017 & 1,050 & 1,043 & 1,017 & 1,050 & 1,043 & 1,017 \\
\textit{R$^2$} & 0.072 & 0.081 & 0.058 & 0.049 & 0.049 & 0.049 & 0.063 & 0.044 & 0.047 \\
\bottomrule\bottomrule
    \end{tabular}%
    \begin{tablenotes}
    \scriptsize \item \textit{Notes}: The table presents results for consumption regularity based on transaction amount, with dependent variables measuring month-to-month volatility in shopping patterns across hours of day (Columns 1–3), days of week (Columns 4–6), and ten-day periods of the month (Columns 7–9). Higher (lower) volatility indicates lower (higher) consumption regularity. \textit{IQ} is the number of correct answers in a seven-question version of Raven’s Progressive Matrices. Dummy for medium (high) education indicates education degree of high school (above high school). Dummy for medium (elder) age refers to those born between 1970 and 1989 (before 1969).  Robust standard errors are in parentheses. {*}~\(p<0.10\), {**}~\(p<0.05\), {***}~\(p<0.01\).
      \end{tablenotes}
    \end{threeparttable}
\end{adjustwidth} 
\end{table}

\newpage

\begin{table}[H]
\begin{adjustwidth}{-1.5cm}{-1.5cm}
\captionsetup{justification=centering}
  \caption{OLS Regressions for Consumption Regularity (Transaction Count)}
  \label{tab:volatility_num}
    \centering
    \footnotesize
\begin{threeparttable}
\begin{tabular}{lccccccccc}
\toprule\toprule
 & \multicolumn{3}{c}{Hours of day} & \multicolumn{3}{c}{Days of week} & \multicolumn{3}{c}{Ten-day periods of the month} \\
\cmidrule(lr){2-4}\cmidrule(lr){5-7}\cmidrule(lr){8-10}
 & 2019 & 2020 & 2021 & 2019 & 2020 & 2021 & 2019 & 2020 & 2021 \\
 & (1) & (2) & (3) & (4) & (5) & (6) & (7) & (8) & (9) \\
\midrule
\textit{Exp: FOSD} & 0.003 & 0.000 & -0.000 & 0.007 & -0.003 & -0.016* & 0.010 & 0.013 & -0.010 \\
\textit{} & (0.003) & (0.004) & (0.004) & (0.007) & (0.009) & (0.009) & (0.010) & (0.015) & (0.017) \\
\textit{Scanner: CCEI} & -0.045*** & -0.039*** & -0.037*** & -0.077*** & -0.077*** & -0.081*** & -0.121*** & -0.118** & -0.107** \\
 & (0.010) & (0.011) & (0.012) & (0.023) & (0.029) & (0.026) & (0.036) & (0.049) & (0.050) \\
\textit{IQ} & -0.000 & -0.000 & -0.000 & 0.000 & -0.001 & -0.001 & 0.001 & -0.000 & 0.001 \\
 & (0.000) & (0.000) & (0.000) & (0.001) & (0.001) & (0.001) & (0.001) & (0.002) & (0.002) \\
\textit{Conscientiousness} & -0.000 & 0.000 & -0.000 & -0.001 & -0.000 & -0.000 & -0.000 & 0.001 & -0.002 \\
\textit{} & (0.000) & (0.000) & (0.000) & (0.001) & (0.001) & (0.001) & (0.001) & (0.002) & (0.002) \\
\textit{Extraversion} & 0.000 & -0.000 & -0.000 & -0.000 & -0.000 & -0.001 & -0.001 & -0.001 & -0.001 \\
\textit{} & (0.000) & (0.000) & (0.000) & (0.001) & (0.001) & (0.001) & (0.001) & (0.002) & (0.002) \\
\textit{Agreeableness} & 0.000 & 0.001* & -0.000 & 0.000 & 0.001 & -0.001 & 0.002 & 0.002 & 0.001 \\
\textit{} & (0.000) & (0.000) & (0.000) & (0.001) & (0.001) & (0.001) & (0.001) & (0.002) & (0.002) \\
\textit{Openness} & 0.001*** & 0.001* & 0.001** & 0.001 & 0.002** & 0.003*** & 0.002* & 0.003 & 0.005*** \\
\textit{} & (0.000) & (0.000) & (0.000) & (0.001) & (0.001) & (0.001) & (0.001) & (0.002) & (0.002) \\
\textit{Neuroticism} & 0.000 & 0.000 & -0.000 & 0.000 & 0.000 & -0.000 & 0.000 & -0.000 & -0.001 \\
\textit{} & (0.000) & (0.000) & (0.000) & (0.001) & (0.001) & (0.001) & (0.001) & (0.002) & (0.002) \\
\textit{Female} & 0.005*** & 0.008*** & 0.007*** & 0.009** & 0.016*** & 0.014*** & 0.019*** & 0.026*** & 0.026*** \\
\textit{} & (0.002) & (0.002) & (0.002) & (0.004) & (0.005) & (0.005) & (0.006) & (0.009) & (0.009) \\
\textit{Medium Income} & 0.000 & -0.002 & -0.001 & 0.000 & -0.003 & -0.002 & 0.004 & -0.009 & 0.006 \\
\textit{} & (0.002) & (0.002) & (0.002) & (0.004) & (0.005) & (0.005) & (0.006) & (0.008) & (0.008) \\
\textit{High income} & 0.003* & -0.001 & -0.001 & 0.007* & -0.001 & -0.001 & 0.012* & -0.003 & 0.001 \\
\textit{} & (0.002) & (0.002) & (0.002) & (0.004) & (0.005) & (0.005) & (0.007) & (0.009) & (0.009) \\
\textit{Medium education} & -0.004** & -0.005*** & -0.005*** & -0.003 & -0.012*** & -0.011** & -0.009 & -0.018** & -0.018** \\
\textit{} & (0.002) & (0.002) & (0.002) & (0.003) & (0.004) & (0.004) & (0.005) & (0.008) & (0.008) \\
\textit{High education} & -0.003 & -0.004* & -0.004* & -0.003 & -0.007 & -0.004 & -0.012* & -0.011 & -0.007 \\
\textit{} & (0.002) & (0.002) & (0.002) & (0.004) & (0.005) & (0.005) & (0.007) & (0.009) & (0.009) \\
\textit{Medium age} & -0.001 & 0.000 & 0.005 & -0.000 & 0.009 & 0.013 & -0.002 & 0.014 & 0.016 \\
\textit{} & (0.003) & (0.003) & (0.003) & (0.007) & (0.008) & (0.009) & (0.012) & (0.013) & (0.016) \\
\textit{Elder age} & -0.010*** & -0.011*** & -0.006* & -0.017** & -0.013 & -0.009 & -0.034*** & -0.021 & -0.021 \\
\textit{} & (0.003) & (0.004) & (0.004) & (0.008) & (0.009) & (0.009) & (0.013) & (0.015) & (0.017) \\
\textit{Family size} & -0.002*** & -0.000 & -0.000 & -0.005*** & -0.002 & 0.001 & -0.007*** & -0.002 & -0.006* \\
 & (0.001) & (0.001) & (0.001) & (0.002) & (0.002) & (0.002) & (0.002) & (0.003) & (0.003) \\
\textit{Constant} & 0.116*** & 0.117*** & 0.120*** & 0.213*** & 0.242*** & 0.265*** & 0.313*** & 0.336*** & 0.389*** \\
 & (0.011) & (0.013) & (0.013) & (0.026) & (0.032) & (0.029) & (0.041) & (0.054) & (0.054) \\
\midrule
N & 1,050 & 1,043 & 1,017 & 1,050 & 1,043 & 1,017 & 1,050 & 1,043 & 1,017 \\
\textit{R$^2$} & 0.079 & 0.083 & 0.068 & 0.058 & 0.054 & 0.056 & 0.071 & 0.049 & 0.052     \\
 \bottomrule\bottomrule
    \end{tabular}%
    \begin{tablenotes}
    \scriptsize \item \textit{Notes}: The table presents results for consumption regularity based on transaction counts, with dependent variables measuring month-to-month volatility in shopping patterns across hours of day (Columns 1–3), days of week (Columns 4–6), and ten-day periods of the month (Columns 7–9). Higher (lower) volatility indicates lower (higher) consumption regularity. \textit{IQ} is the number of correct answers in a seven-question version of Raven’s Progressive Matrices. Dummy for medium (high) education indicates education degree of high school (above high school). Dummy for medium (elder) age refers to those born between 1970 and 1989 (before 1969).  Robust standard errors are in parentheses. {*}~\(p<0.10\), {**}~\(p<0.05\), {***}~\(p<0.01\).
      \end{tablenotes}
    \end{threeparttable}
\end{adjustwidth} 
\end{table}

\newpage

\section{Economic Specifications and Estimations of Preferences}
\label{app:est}

Given any period, a DM's choices are denoted as $(x_1, x_2)$. For risk preference, $x_1$  ($ x_2$) corresponds to the better (worse) outcome of contingent assets, respectively. Regarding social preference, $x_1$ ($x_2$) represents her own (opponent's) payment. For food preference, they are the consumption quantities on two types of food. For risk preference, we apply the commonly used disappointment aversion (DA) model, which is equivalent to rank-dependent utility when there are two equiprobable states, specified in Equation \ref{eq-risk}; while for social and food preferences, we use the constant elasticity of substitution (CES) utility function \citep{fisman2007individual} specified in Equation~\ref{eq-food-social}. 
\begin{equation}\label{eq-risk}
    U(x_1,x_2)=\alpha u(x_1)+(1-\alpha)u(x_2)\text{, } \alpha \in [0,1]\text{, } 
    u(z)= \begin{cases} \frac{1}{\rho}z^\rho,\rho\le 1(\rho \ne 0)\\ \ln(z), \rho=0 \end{cases}
\end{equation}

\begin{equation}\label{eq-food-social}
    U(x_1,x_2)=[\alpha x_1^\rho +(1-\alpha)x_2^\rho]^{\frac{1}{\rho}}\text{, } \alpha \in [0,1]\text{, } \rho \leq 1
\end{equation}

Specifically, $\alpha$ captures the relative weight on better outcome in risk task, the degree of selfishness for social choices, and the weight on the first commodity in food purchase. $\rho$ indicates the degree of risk-seeking in risk preference. For social and food preferences, $\rho$ captures the curvature of the indifference curves. As $\rho$ approaches 1, the two accounts (goods) become perfect substitutes. Conversely, as $\rho$ approaches negative infinity, the two accounts (goods) become perfect complements. 

Referring to the method by \citet{fisman2007individual}, the demand function for the two utility models is given by
$$x_1=\left[\frac{g}{(p_1/p_2)^m+g} \right] \frac{E}{p_1}, $$
where $E$ is the expenditure, $ m=\rho/(1-\rho)$, and $g=\left[\alpha/(1-\alpha)\right]^{1/(1-\rho)}$. This generates the following econometric specification for estimation:$$\frac{p_1^t x_1^t}{E^t}=\frac{g}{(p_1^t/p_2^t)^{m}+g}+\varepsilon^t,$$ where $\varepsilon^t$ is assumed to be normally distributed with mean zero and variance $\sigma^t$. We generate the estimates $\hat{g}$ and $\hat{m}$, using the nonlinear Tobit maximum likelihood method, then use these to calculate  parameters $\hat{\alpha}$ and $\hat{\rho}$, and regard log likelihood (\textit{LL}) as a measure of goodness of fit.

\newpage
\section{Experimental Instructions}
\label{instructions}

\setcounter{table}{0}
\renewcommand{\thetable}{C\arabic{table}}
\setcounter{figure}{0}
\renewcommand{\thefigure}{C\arabic{figure}}

\subsection{Risk Task}\label{instru_risk}

As your choices may influence the bonus you receive, please ensure that you understand the following instructions and answer carefully.

\begin{itemize}

\item This experiment consists of 22 rounds. In every round, you have 100 tokens, which can be allocated between the blue membership account (Blue Account) and red membership account (Red Account).

\item In the following example:
\begin{itemize}
    \item Blue Account: 1 token = RMB 0.8, that is, each token for the Blue Account is worth RMB 0.8
    \item Red Account: 1 token = RMB 0.2, that is, each token for the Red Account is worth RMB 0.2
\end{itemize}

\begin{figure}[H]
\centering
\includegraphics[width=0.65\textwidth]{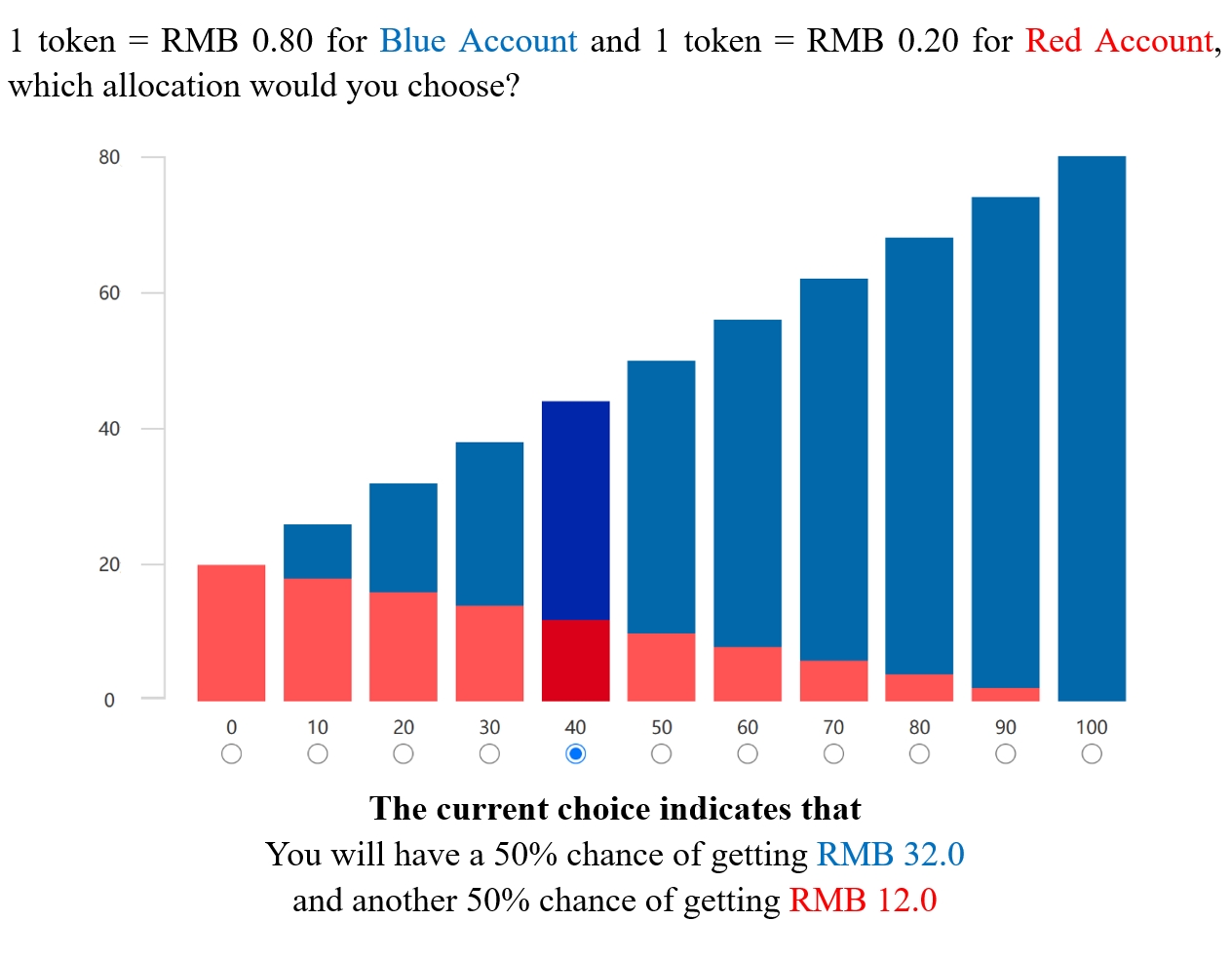}
\end{figure}

\item As shown in the figure, you need to create an allocation plan. Different allocation plans influence the amount of money in your accounts, that is,

\begin{itemize}
    \item if 0\% of the tokens are allocated to the Blue Account and 100\% to the Red Account, then there will be RMB 0 in the Blue Account and RMB 20 in the Red Account;
    \item if 10\% of the tokens are allocated to the Blue Account and 90\% to the Red Account, then there will be RMB 8 in the Blue Account and RMB 18 in the Red Account;
    \item ...
    \item if 100\% of the tokens are allocated to the Blue Account and 0\% to the Red Account, then there will be RMB 80 in the Blue Account and RMB 0 in the Red Account.
\end{itemize}

\item In each decision-making round, you have a 50\% chance of getting the money in the Blue Account and another 50\%  chance of getting the money in the Red Account. As shown in the figure, when you choose 40 (40\% of the tokens are allocate to the Blue Account and 60\% to the Red Account), you will find the following note below the bar graph, ``You will have a 50\% chance of getting RMB 32.0 and another 50\% chance of getting RMB 12.0''. When you change the option, the amounts of money in the note will change accordingly.

\item Different choices will lead to different gains and risks.
\begin{itemize}
    \item The gain is reflected in the total length of the two-colored bars: the longer the bar is, the greater the gain is; and the shorter the bar is, the smaller the gain is.
    \item The risk is reflected in the difference between the length of the two colors within the bar: the greater the difference is, the higher the risk is; and the smaller the difference is, the lower the risk is.
\end{itemize}

\item At the end of the experiment, the computer will select one decision round randomly, where each account has an equal probability of being chosen, and the participant will be paid the amount he/she has earned in that round.
\end{itemize}

\subsection{Social Task}\label{instru_social}

As your choices may influence the bonus you receive, please ensure that you understand the following instructions and answer carefully.
    \begin{itemize}
    \item This experiment consists of 22 rounds. In every round, you have 100 tokens, which can be allocated between yourself and another supermarket consumer. 
    \item In the following example:
    \begin{itemize}
        \item 1 token = RMB 0.4 for the other
        \item 1 token = RMB 0.2 for yourself
        \end{itemize}

\begin{figure}[H]
\centering
\includegraphics[width=0.65\textwidth]{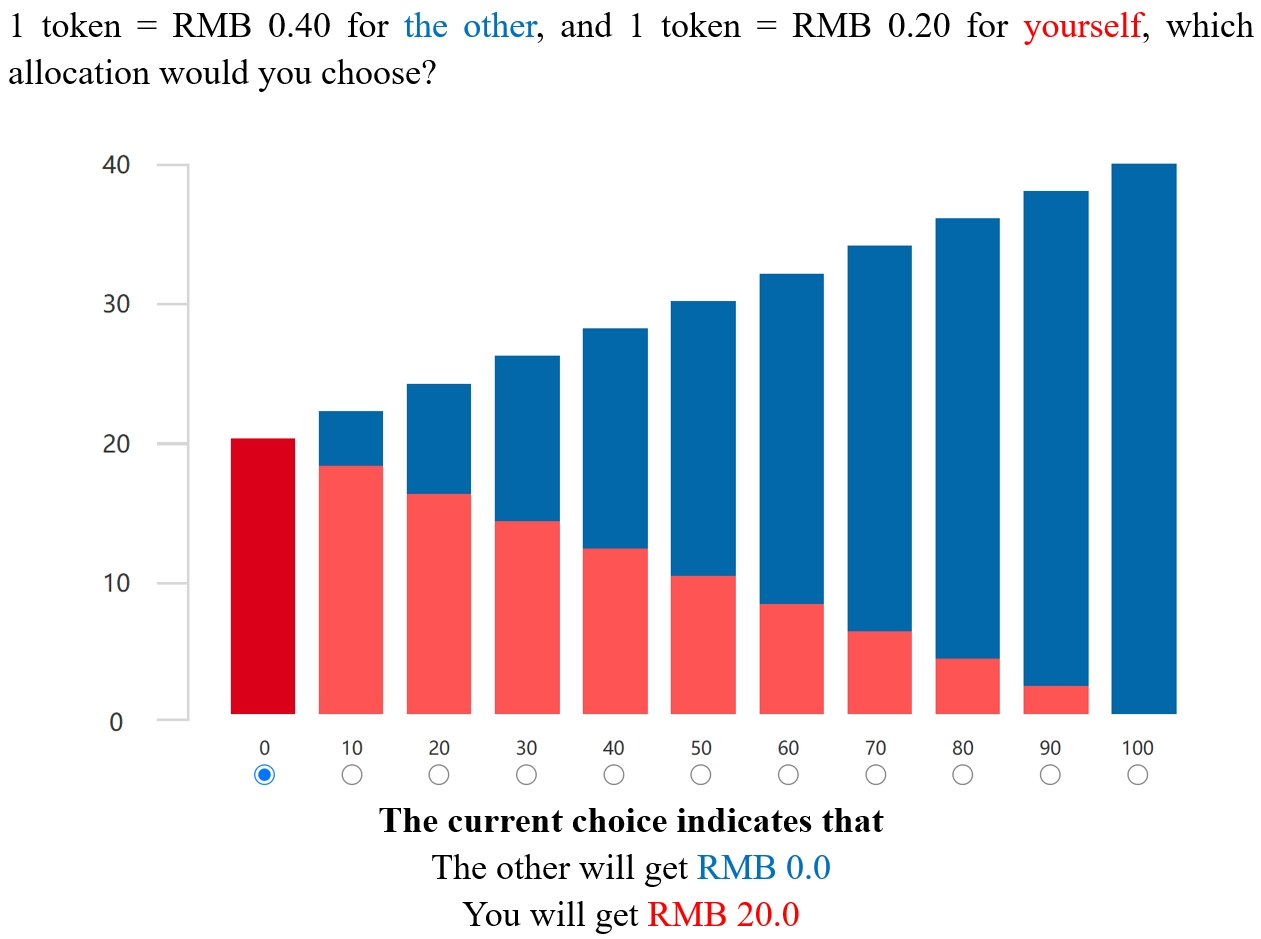}
\end{figure}

   \item As shown in the figure, you need to create an allocation plan. Different allocation plans influence the amount of money between yourself and the other, that is,    
      
    \begin{itemize}
        \item if 0\% of the tokens are allocated to the other and 100\% to yourself, then the other will have RMB 0, and you will have RMB 20;
        \item if 10\% of the tokens are allocated to the other and 90\% to yourself, then the other will have RMB 4, and you will have RMB 18;
        \item ...
        \item if 100\% of the tokens are allocated to the other and 0\% to yourself, then the other will have RMB 40, and you will have RMB 0.
    \end{itemize}

\item As shown in the figure, when you choose 0 (0 \% of the tokens are allocate to the other and 100\% to yourself), you will find the following note below the bar graph, ``The other will get RMB 0.0, and you will get RMB 20.0''. When you change the option, the amounts of money in the note will change accordingly.

\item At the end of the experiment, the computer will select one decision round randomly, and the participant and another randomly matched participant will be paid the amounts they have earned in that round.
\end{itemize}

\subsection{Food Task}\label{instru_food}

As your choices may influence the bonus you receive, please ensure that you understand the following instructions and answer carefully. 
\begin{itemize}
    
    \item This experiment consists of 22 rounds. In every round, you have RMB 50, which can be allocated to buy tomatoes and hams.
    
    \item In the following example:
    \begin{itemize}
        \item tomato: RMB 10/kg
        \item picnic ham: RMB 20/kg
        
    \end{itemize}

\begin{figure}[H]
\centering
\includegraphics[width=0.5\textwidth]{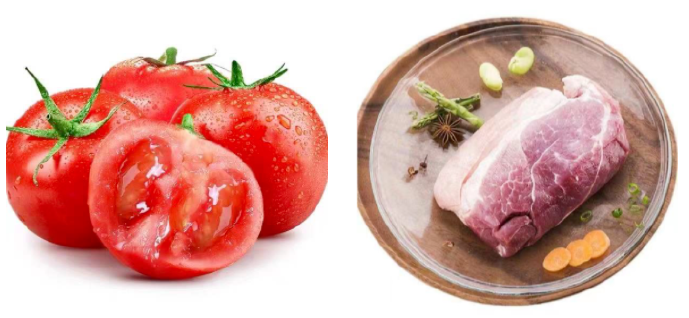}
\end{figure}

    \item As shown in the figure, you need to create an allocation plan. Different allocation plans influence the amount of picnic hams and tomatoes you buy, that is,    
      
    \begin{itemize}
        \item if 0\% of the expenditure is allocated to hams and 100\% to tomatoes, then you will buy 0~kg of picnic hams and 5~kg of tomatoes;
        \item if 10\% of the expenditure is allocated to hams and 90\% to tomatoes, then you will buy 0.25~kg of hams and 4.5~kg of tomatoes;
        \item ...
        \item if 100\% of expenditure is allocated to hams and 0\% to tomato, then you will buy 2.5~kg of hams and 0~kg of tomatoes. 
    \end{itemize}

\begin{figure}[H]
\label{fig-insructions1} 
\centering
\includegraphics[width=0.65\textwidth]{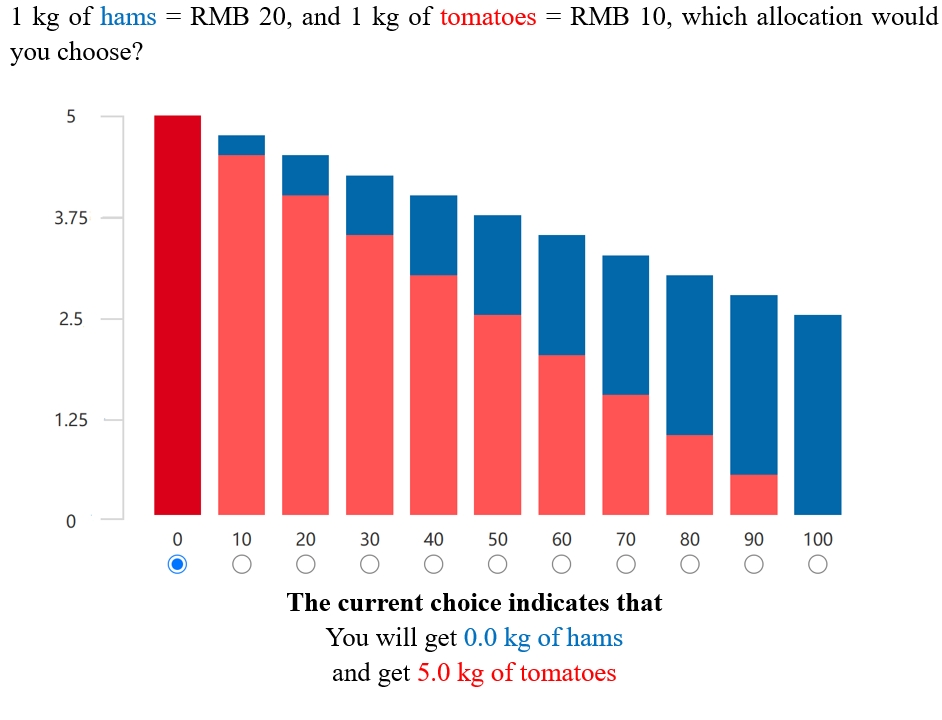}
\end{figure}

\item As shown in the figure, when you choose 0 (0\% of the expenditure is allocated to hams and 100\% to tomatoes), you will find the following note below the bar graph: ``You will get 0.0~kg of hams and 5.0~kg of tomatoes''. When you change the option, the amounts of products in the note will change accordingly.

\item At the end of the experiment, the computer will select one decision round randomly, and the vouchers of products the participant has earned in that round will be deposited into his/her membership account.

\end{itemize}

\clearpage

\subsection{Post-Experiment Questionnaire}\label{instru_post}
\begin{itemize}

    \item Big Five Personality Traits Test\\
(7 Likert Scale) I see myself as someone who
\begin{itemize}
    \item is reserved
    \item  is generally trusting
    \item  tends to be lazy
    \item is relaxed, handles stress well
    \item has few artistic interests
    \item is outgoing, sociable
    \item tends to find fault with others
    \item does a thorough job
    \item gets nervous easily
    \item has an active imagination
    
\end{itemize}

\item Raven's IQ Test\\
In each of the following questions, a part of the graph is missing from its lower right side,  please find the appropriate graph to fill in the gap. There is only one correct answer for each question. 

\begin{figure}[H]

\centering
\includegraphics[width=\textwidth]{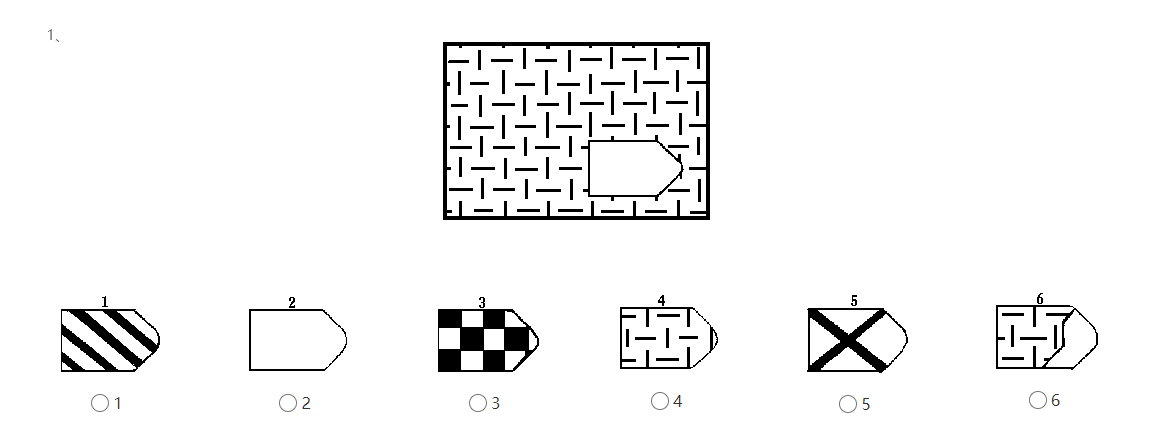}
\end{figure}

\item Demographics 
\begin{itemize}
\item Gender:
\item Year of Birth:
\item Number of household members in your household:
\item Your Hukou (residency status) is: Urban/Rural
\item Individual monthly income:
\item Household monthly income:
\item Your education level:
\end{itemize}

\end{itemize}

\end{document}